\documentclass[11pt,a4paper]{article}
\usepackage[margin=22mm]{geometry}
\usepackage{fontspec}
\usepackage{amsmath,amssymb,bm,graphicx,array,booktabs}
\usepackage{needspace,xurl}
\usepackage[bookmarks=true,unicode=true,hidelinks]{hyperref}
\hypersetup{pdftitle={Localization of Candidate Kikuchi Regions in RHEED Images: Visibility and Annotation Boundaries},
  pdfauthor={Lumou Weng, Hanshan Huang, Zhenhan Zhang, Gan Wang}}
\title{Localization of Candidate Kikuchi Regions in RHEED Images: Visibility and Annotation Boundaries}
\author{Lumou Weng, Hanshan Huang, Zhenhan Zhang, Gan Wang\thanks{Corresponding author: \href{mailto:wangg@sustech.edu.cn}{wangg@sustech.edu.cn}}\\[3mm]
  \small Department of Physics, Southern University of Science and Technology\\
  \small Shenzhen 518055, China}
\date{}
\begin{document}
\maketitle
\section*{Abstract}

\addcontentsline{toc}{section}{Abstract}

Kikuchi lines and bands in reflection high-energy electron diffraction (RHEED) carry information on crystal geometry and electron scattering, but are often obscured by intense diffraction streaks. We combine multiscale convolution with spatial detail from skip connections and independent supervision that allows overlapping regions to localize streaks and candidate Kikuchi regions separately. Using manual polygon annotations made before model-assisted editing, the Kikuchi intersection-over-union (IoU) on 29 laboratory test images from separate growth batches was 0.6290±0.0137. We then performed supervised adaptation to public chalcogenide images and stratified images by Kikuchi visibility using the median ratings of three observers who rated the images separately, two of whom rated with predictions hidden. With all 244 training images, IoU for the clear-feature group of 10 test images was 0.5014±0.0336, whereas ambiguous images gave lower values. With 100 training images, clear-group IoU was 0.5171±0.0073; expanding the training set also reduced responses on images without target features. Reported uncertainties are sample standard deviations across four laboratory runs or three adaptation runs. Public-image adaptation used reference masks of mixed provenance, including prediction-derived drafts. The method converts visual cues into inspectable spatial regions, providing a basis for analysis of line positions, intersections, and local intensity.\par

Keywords: molecular beam epitaxy; reflection high-energy electron diffraction; Kikuchi structures; computer vision; region localization; visibility grading\par

\section*{1. Introduction}

\addcontentsline{toc}{section}{1. Introduction}

Reflection high-energy electron diffraction (RHEED) is widely used for in situ characterization during molecular beam epitaxy (MBE), allowing diffraction streaks, spots, surface reconstructions, and intensity changes to be followed during growth [17]. In addition to these commonly used features, patterns can contain Kikuchi lines and bands. Previous studies, for example, observed Kikuchi lines on FeTe surfaces in the \(\mathrm{Bi_2Te_3/FeTe}\) system and on \(\mathrm{Al_2O_3}\) surfaces used for epitaxy [26,27]. This study addresses how to extract these structures separately under different growth and imaging conditions, across material systems.\par

Kikuchi structures contain physical information that merits further use. The angular distribution of diffusely scattered electrons undergoes subsequent diffraction in the crystal, producing lines and bands related to crystallographic plane geometry. Under the grazing-incidence conditions of RHEED, refraction and surface-wave resonance also affect line shape and local contrast [1,2]. RHEED Kikuchi patterns have been used to identify crystal axes and analyze surface relaxation and electron propagation conditions [3–6]. Electron backscatter diffraction (EBSD) has developed Kikuchi-pattern analysis for orientation and local lattice-deformation measurements [9–11]. These applications each require geometric calibration and an appropriate scattering description, but all first require the structures and their positions to be identified. In RHEED, overlap between low-contrast Kikuchi bands and intense streaks makes this step difficult.\par

Automated RHEED analysis has expanded from intensity tracking to whole-pattern and local-structure analysis. Dimensionality reduction and clustering represent patterns with a small set of descriptors and organize growth processes by similarity [12,13]. Automated structural phase mapping and reconstruction of azimuthal patterns acquired by rotating about the surface normal further extend the range of analysis [14,16]. Kaspar et al.'s RHAAPsody workflow demonstrates frame-by-frame analysis of growth sequences [23]. Muetzel et al. trained convolutional networks to classify overall chalcogenide RHEED pattern types and released their images and code [18,19]; the Streaks category provides a reusable external source for studying local structures. Price et al. combined U-Net with transformer models that relate information across image positions for pixel-wise segmentation, using diffraction features to predict grain orientation and composition [15]. These studies demonstrate the value of automated analysis for materials growth, while complete localization of weaker Kikuchi bands remains an important challenge.\par

This work arose from observations during MBE growth in our laboratory: bright streaks were readily extracted, but weak Kikuchi structures were often overlooked because of obscuration, discontinuity, and changes in orientation. Digital filtering can improve their visibility [30], whereas producing separate region outputs also requires distinguishing bright structures from target structures. Li et al.'s LarkMNet study of weak cracks against prominent construction joints [29] provided an important inspiration: weak-feature recognition benefits from preserving local detail while using morphological relationships over a wider area. We implemented this principle in a U-Net-style pixel-segmentation network with multiscale encoding and reconstruction, supervising streak and Kikuchi regions separately so that both can be retained at intersections.\par

We refer to the targets as “candidate Kikuchi regions,” encompassing visible Kikuchi lines and bands and features with recognizable Kikuchi-like morphology. Known polycrystalline rings, central diffraction rings, and bright screen edges are non-targets. This terminology distinguishes image localization from crystallographic indexing: region recognition identifies positions to inspect, while physical assignment requires diffraction geometry and experimental context. Figure 1 illustrates the parallel streak and Kikuchi outputs. We first evaluate localization using the laboratory's original manual annotations, then examine the additional training data needed for adaptation to public images. Independent visibility ratings and region re-annotation help explain performance differences between images.\par

\begin{center}\begin{minipage}{\linewidth}\centering
\includegraphics[width=\linewidth,height=0.62\textheight,keepaspectratio]{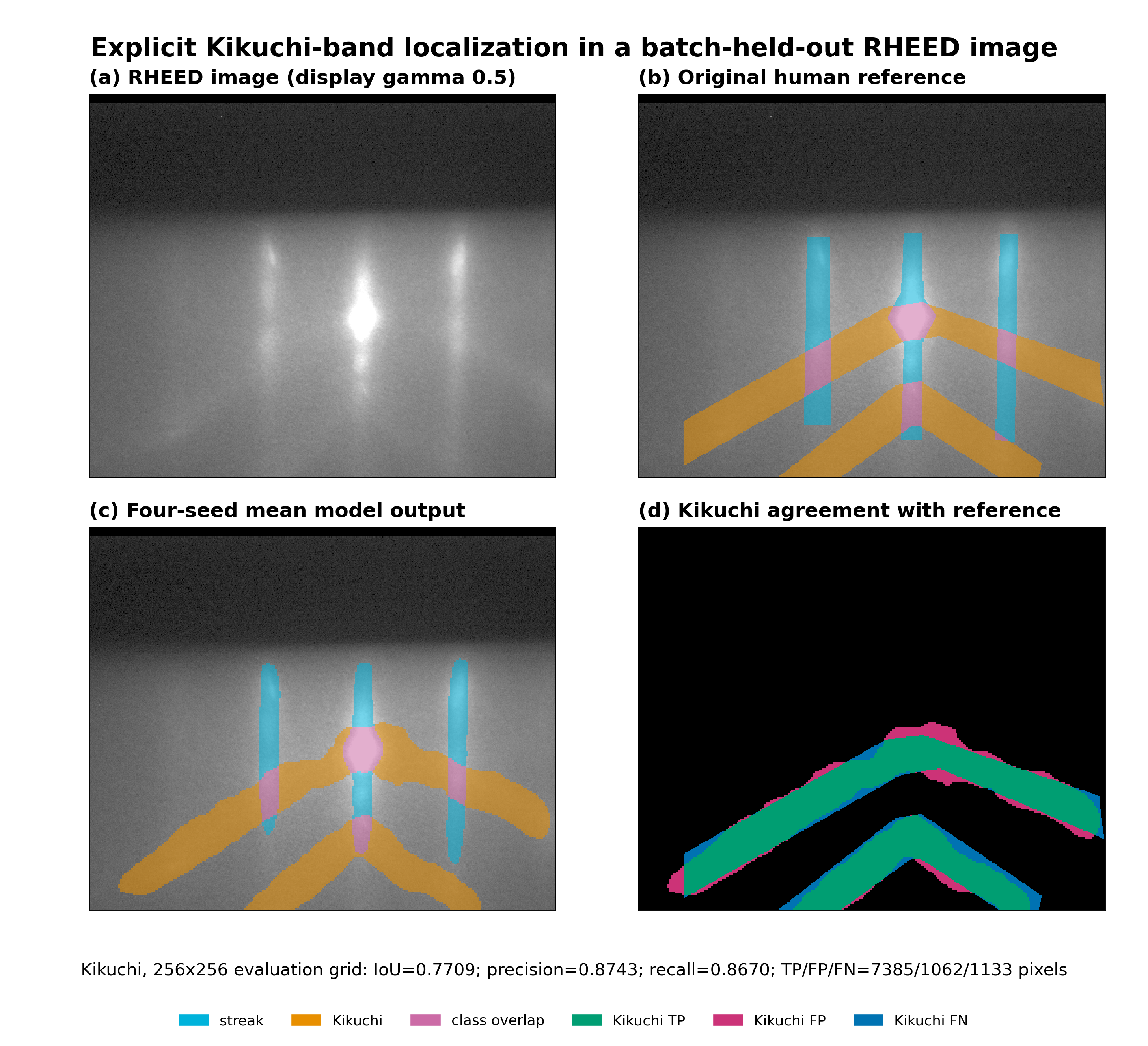}
\par\vspace{3mm}{\small\raggedright Figure 1. Localization of streaks and candidate Kikuchi regions in a laboratory test image. (a) Original image; (b) original manual reference made before model-assisted editing; (c) predictions obtained by thresholding the mean class scores from four independent training runs at 0.45; (d) Kikuchi-region errors. Reference and prediction overlays use cyan for streaks, yellow for Kikuchi regions, and pink for overlap; the error map uses green for correctly covered pixels, magenta for extra predictions, and blue for omissions. For display, normalized brightness \(x\in[0,1]\) is transformed to \(\sqrt{x}\) to reveal dark features; this operation does not enter the model or metric calculation. This clear illustration was chosen from high-scoring candidates from each test batch and has Kikuchi IoU 0.7709. Overall results are given in Table 2; the full distribution and fixed low-, middle-, and high-IoU examples appear in Figures S2 and S12. Instrument text cropped from the display contains no reference or predicted foreground; metrics use the complete 256×256 grid.\par}
\end{minipage}\end{center}

\section*{2. Image Representation and Localization Method}

\addcontentsline{toc}{section}{2. Image Representation and Localization Method}

\subsection*{2.1 Grayscale Images and Overlapping Regions}

\addcontentsline{toc}{subsection}{2.1 Grayscale Images and Overlapping Regions}

A digital image consists of pixels arranged on a regular grid; a grayscale image stores one brightness value at each pixel. The method uses a 256×256 grayscale input, with values normalized to 0–1 when read. Recorded brightness depends on physical diffraction intensity as well as exposure, gain, and phosphor-screen condition. A manual region annotation is stored as a separate position map aligned with the image, called a mask: 1 marks a target location and 0 a non-target location.\par

Streak and Kikuchi structures can intersect in the projected pattern. We therefore store separate binary references \(G_s(x)\) and \(G_k(x)\), allowing both to equal 1 at the same pixel \(x\). Their union \(G_u(x)=\max[G_s(x),G_k(x)]\) indicates the presence of at least one target, while their intersection retains locations shared by both structures. Image-level visibility grades and pixel-wise masks are recorded separately: the former describe how clearly Kikuchi features can be discerned, and the latter define the spatial reference used for localization.\par

\subsection*{2.2 Separate Streak and Kikuchi Localization}

\addcontentsline{toc}{subsection}{2.2 Separate Streak and Kikuchi Localization}

The model learns relationships between local brightness variations and line or band morphology from annotated examples. Its basic organization follows U-Net-style encoding and reconstruction [22]. Convolution moves a small window across an image, combining neighboring values with the same set of learned coefficients. Multiple coefficient sets produce multiple response maps, each called a feature channel. Each convolution block contains two 3×3 convolutions, each followed by ReLU, which sets negative responses to zero and retains positive responses so that multiple layers can express more complex pattern relationships. These learned features describe image morphology and brightness distributions; crystallographic assignments and physical quantities require further diffraction-geometric analysis.\par

The encoder retains 48, 96, and 192 channels on 256×256, 128×128, and 64×64 grids, respectively (Figure 2). Two 2×2 max-pooling operations each halve both grid dimensions, combining information over larger original-image regions on smaller grids. Reconstruction restores the grid by bilinear interpolation and concatenates encoder responses from the same scale along the channel dimension, forming skip connections. Wider structural relationships and fine spatial detail from shallower layers thus both contribute to localization of elongated regions.\par

A separate branch at the smallest 64×64 grid relates different image regions. It first aggregates spatial information onto an 8×8 grid, then learns associations between positions so that each position combines responses from all positions according to their association weights. This operation is called self-attention. The result is restored to 64×64, multiplied by a trainable coefficient \(\gamma\), and added to the original responses; \(\gamma\) is initialized to 0. Exact dimensions, numerical operations, and random channel suppression during training are given in Section S2, with controls in S8. Section S11 explains the terminology for physics readers.\par

\begin{center}\begin{minipage}{\linewidth}\centering
\includegraphics[width=\linewidth,height=0.62\textheight,keepaspectratio]{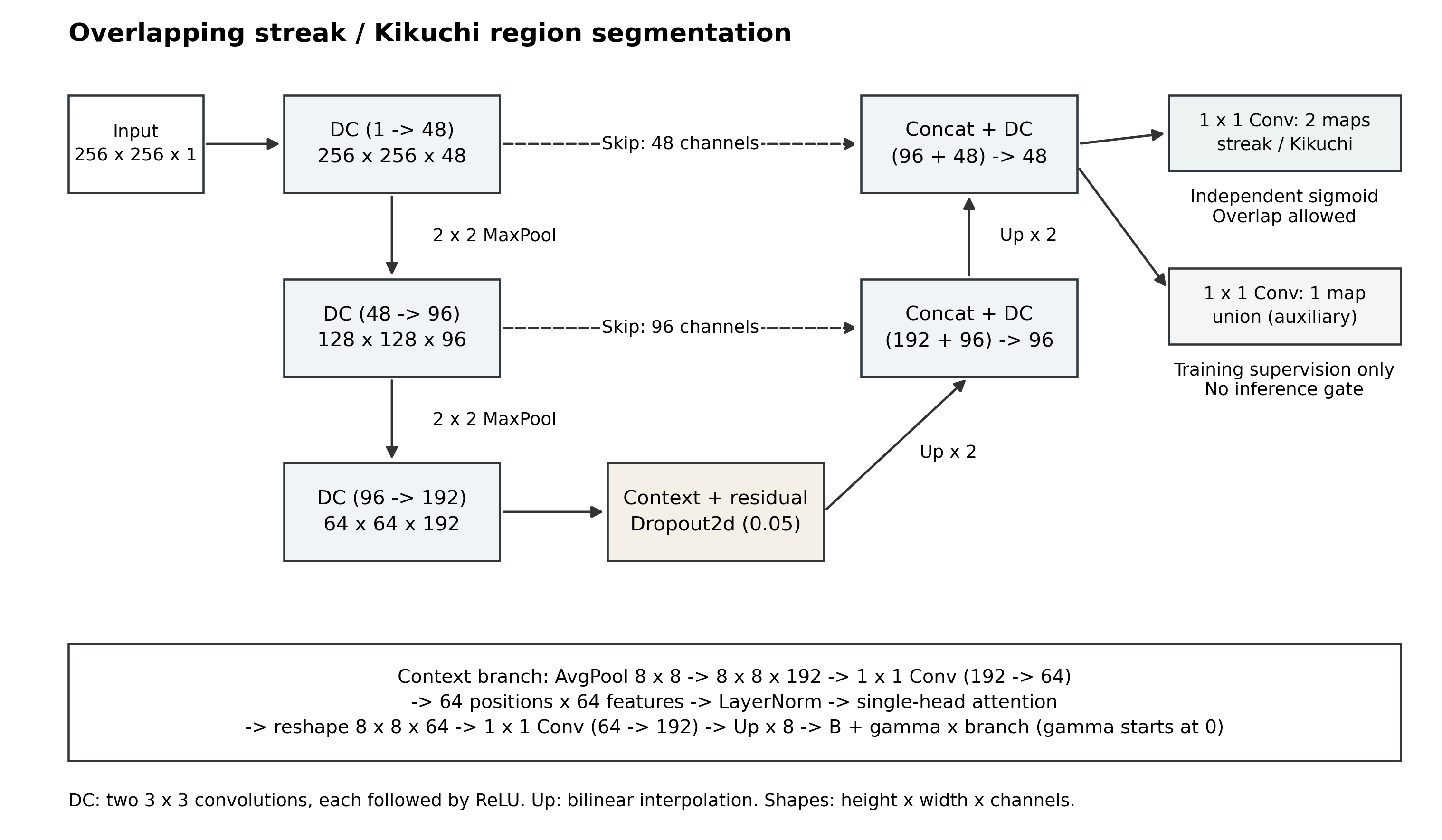}
\par\vspace{3mm}{\small\raggedright Figure 2. Network for separate streak and Kikuchi localization. DC denotes two 3×3 convolutions, each followed by ReLU; MaxPool denotes 2×2 max pooling; Up denotes bilinear upsampling. Dashed lines show skip concatenations between encoder and reconstruction paths. Grids are labeled as height×width×channels. The main path has only two pooling steps; the compressed-context branch uses additional average pooling. Final 1×1 convolutions produce an auxiliary union output and two class outputs for streaks and Kikuchi regions. The complete context-branch sequence is shown below for comparison with the text and code. The base network contains 1,100,676 trainable parameters, including 41,601 in the context branch.\par}
\end{minipage}\end{center}

The outputs pass through the sigmoid function \(p=1/(1+e^{-z})\), mapping a real value \(z\), called a logit, to a score \(p\) between 0 and 1 for that class. Streak and Kikuchi score maps are thresholded separately, allowing intersections to belong to both classes. The union output provides auxiliary supervision during training; prediction uses the two class outputs directly. Computing outputs for a new image with fixed parameters is called inference, as distinct from training, which adjusts the parameters. Scores and thresholds are explained in S11.5, and complete pseudocode for preprocessing, network calculation, training, and evaluation appears in S12.\par

\subsection*{2.3 Supervision of Weak Structures}

\addcontentsline{toc}{subsection}{2.3 Supervision of Weak Structures}

A loss function expresses prediction–reference discrepancies as a numerical objective for updating parameters. Even when Kikuchi pixels are darker than streak pixels, omitting them incurs a training penalty if the reference marks them as targets; predicting Kikuchi regions in the background is also penalized. Binary cross-entropy (BCE) is computed independently over the whole image for each class:\par

\begin{equation*}
\begin{aligned}
L_{\mathrm{BCE},c}
&=-\frac{1}{N}\sum_x \Bigl[
G_c(x)\log p_c(x)\\
&\hspace{4em}+[1-G_c(x)]\log[1-p_c(x)]\Bigr].
\end{aligned}
\end{equation*}

Here \(N\) is the number of evaluated pixels, and \(c=u,s,k\) denotes union, streak, and Kikuchi, respectively. A reference value of 1 with a score near 0, or a reference value of 0 with a score near 1, produces a large loss. Both positive requirements remain active at overlapping pixels. This supervision reflects the projected pattern geometry more closely than a representation that assigns only one class to an intersection.\par

The total loss for the primary internal experiment is\par

\begin{equation*}
L=L_{\mathrm{BCE},u}+\tfrac12\bigl(L_{\mathrm{BCE},s}+L_{\mathrm{BCE},k}\bigr)+0.2L_{D,k},
\end{equation*}

\begin{equation*}
L_{D,k}=1-\frac{2\sum_xp_k(x)G_k(x)+1}{\sum_xp_k(x)+\sum_xG_k(x)+1}.
\end{equation*}

The soft-Dice term \(L_{D,k}\) emphasizes overlap of the total Kikuchi region using continuous scores. “Soft” means that scores are not first converted to 0 or 1, allowing continuous calculation of the parameter-update direction. The added 1 stabilizes calculations for empty regions, and 0.2 is the weighting coefficient. Independent BCE already produced Kikuchi localization during development before this term was introduced; the current loss adds a region-overlap requirement to that foundation (S2.3). Public-image adaptation uses the same localization network with additional image-visibility supervision (Section 3.3).\par

\subsection*{2.4 Localization and False-Positive Metrics}

\addcontentsline{toc}{subsection}{2.4 Localization and False-Positive Metrics}

Pixels covered by both reference \(G\) and prediction \(P\) are true positives (\(TP\)); pixels covered only by the prediction are false positives (\(FP\)), and pixels covered only by the reference are false negatives (\(FN\)). Intersection-over-union is defined as\par

\begin{equation*}
\mathrm{IoU}=\frac{|P\cap G|}{|P\cup G|}=\frac{TP}{TP+FP+FN}.
\end{equation*}

It measures the overlapping area as a fraction of the total area covered by either mask and is affected by omissions, extra predictions, position, and boundary width. Precision, \(TP/(TP+FP)\), measures how much of the predicted region is supported by the reference; recall, \(TP/(TP+FN)\), measures how much of the reference is recovered. F1, equal to Dice for these binary regions, is \(2TP/(2TP+FP+FN)\). Elongated structures are particularly sensitive to transverse displacement: for two long strips of equal width \(w\) displaced by \(\delta<w\), IoU is \((w-\delta)/(w+\delta)\) when end effects are neglected.\par

The primary metric first sums \(TP/FP/FN\) over an image set for each model and then computes IoU, termed pooled pixel-level IoU. We then report the mean and sample standard deviation across independent training runs. Additional internal statistics first average the four model IoUs for each image, then average with equal weight per image. Equal-batch weighting first averages images within each batch and then averages the six batches. When both reference and prediction are empty, per-image IoU is 0/0 and is reported as not applicable (NA); false positives on empty-reference images remain in the denominator of overall IoU. For images without targets, we also report \(FP\) as a fraction of valid image area to show directly how much area is erroneously marked. Training weights based on visibility grades do not weight test metrics.\par

\section*{3. Experiments}

\addcontentsline{toc}{section}{3. Experiments}

\subsection*{3.1 Image Sources, Annotations, and Data Partitions}

\addcontentsline{toc}{subsection}{3.1 Image Sources, Annotations, and Data Partitions}

Our group uses MBE to study layered chalcogenides, magnetic tellurides, and their heterostructures. Previous studies used elemental or compound sources to grow films on oxide and semiconductor substrates [24,25,28]. RHEED images were acquired with an R-DEC RDA-004P system (Japan) at an accelerating voltage of 15 kV and a typical beam incidence angle of approximately 5° relative to the sample surface. Archives cover sample loading, degassing, before, during, and after growth, and annealing, but experimenters preferentially retained images with well-defined patterns. Exposure and gain were adjusted for observation; deposits, mainly Te, progressively dimmed the phosphor screen and required higher gain. The historical images therefore represent selected records acquired under varying conditions.\par

The primary internal dataset contains 145 images of 512×512 pixels, preferentially selected for visible Kikuchi structures. Original polygons define 592 streak regions and 381 Kikuchi regions. We identified 34 growth batches from filenames, original-image correspondence, folder structure, and modification timestamps. Training, validation, and test partitions keep complete batches separate (Table 1), preventing similar images from the same growth run from entering different partitions. All primary internal results use original manual polygons completed before model-prediction overlays and adoption functions were enabled, without revisions based on model outputs.\par

Public images come from Muetzel et al.'s chalcogenide RHEED dataset [18,19]. Its Streaks category provides streak-pattern backgrounds different from the laboratory images; we added two-class region references and image-level visibility ratings. Of 382 images rated by three observers, one lacked reliable sample-group information and was excluded before partitioning. The remaining 381 images were separated by sample group identified from file information. Public visibility strata in Table 1 use the same test results, rather than constituting additional independent experiments.\par

\par\noindent\begin{minipage}{\linewidth}\small
Table 1. Data partitions for the two experiments. Counts refer to original images before augmentation. “Nonempty/\allowbreak{}empty” in the public test set refers to the fixed Kikuchi reference masks.\par\vspace{4pt}
\setlength{\tabcolsep}{3pt}
\renewcommand{\arraystretch}{1.2}
\begin{tabular}{@{}>{\raggedright\arraybackslash}p{\dimexpr 0.296774\linewidth-2\tabcolsep\relax}>{\raggedright\arraybackslash}p{\dimexpr 0.206452\linewidth-2\tabcolsep\relax}>{\raggedright\arraybackslash}p{\dimexpr 0.116129\linewidth-2\tabcolsep\relax}>{\raggedright\arraybackslash}p{\dimexpr 0.083871\linewidth-2\tabcolsep\relax}>{\raggedright\arraybackslash}p{\dimexpr 0.296774\linewidth-2\tabcolsep\relax}@{}}
\toprule
\textbf{Experiment} & \textbf{Training images} & \textbf{Validation images} & \textbf{Test images} & \textbf{Source separation and test composition} \\
\midrule
Laboratory localization with original annotations & 93 & 23 & 29 & 34 batches overall; 6 test batches excluded from training and validation; all 29 references nonempty \\
Public Streaks adaptation and stratification & 20 to 220 in steps of 20, or 244 & 51 & 86 & 38/\allowbreak{}11/\allowbreak{}12 training/validation/test groups; 34 nonempty and 52 empty test references \\
\bottomrule
\end{tabular}
\end{minipage}\par

Public reference masks combine three sources: 203 earlier manual annotations, 40 prediction drafts with manual review, and 138 prediction-initialized drafts unreviewed in the fixed provenance records. Some images and historical outputs had also been used during method development. The public experiment therefore measures retrospective supervised adaptation against these mixed-provenance references, rather than prospective blinded performance against independently drawn pixel-level ground truth. This distinction applies throughout the public learning curves and stratified statistics; itemized provenance and counts are provided in S1.\par

Internal images are read as 8-bit grayscale, divided by 255, and resized to 256×256. Public images retain their original bit depth: 8-bit and 16-bit grayscale values are divided by 255 and 65535, respectively, resized while preserving aspect ratio, and centered with black padding to the same size. Normalization uses the maximum specified by the data type, rather than the brightest value in each image. Padding contributes as background during public training. Evaluation first thresholds predictions at 0.45 on the 256×256 grid, removes padding, and restores binary masks to the original image dimensions by nearest-neighbor interpolation for comparison with original-size references. Internal evaluation uses the 256×256 grid directly. The two experiments therefore have their own references and evaluation grids. Detailed interpolation rules are given in S1.3 and Algorithms S1 and S5. Neither primary experiment uses the earlier +30 brightness augmentation or fusion of original and brightened inputs; square-root brightening in Figures 1 and 3 is solely for display.\par

\subsection*{3.2 Laboratory Training and Controls}

\addcontentsline{toc}{subsection}{3.2 Laboratory Training and Controls}

Training adjusts parameters using loss gradients, which describe how the loss changes with each parameter. One adjustment is called a parameter update; a complete pass through the training views is an epoch. We process one image at a time with AdamW: the learning rate determines the update step size, and weight decay limits excessive parameter growth. Four runs start from independent random parameters, with seeds 42, 7, 123, and 2026. The initial learning rate is \(8\times10^{-5}\), weight decay is \(10^{-4}\), and the maximum is 80 epochs. Plateaus in validation loss control learning-rate reduction, whereas checkpoints are saved on a new highest validation pooled Kikuchi F1 at threshold 0.50. Ties retain the earlier model; training stops after 25 consecutive epochs without a new best F1. Test evaluation uses the prespecified threshold 0.45 and does not determine the learning rate, checkpoint, or threshold. Epoch-level curves and complete settings appear in S3 and Algorithm S3.\par

Training images and masks undergo synchronized D4 augmentation: rotations by multiples of 90° and their reflections produce eight views. All eight views per image enter the training list, which is shuffled and traversed; 93 original images thus provide 744 views and updates per epoch, rather than one randomly chosen transformation per image. Validation and test images retain their original orientation, and image counts refer to originals. Six paired controls examine the context branch, skip concatenations, output parameterization, union supervision, soft-Dice, and D4. The D4 control repeats each original image eight times to match views per epoch and the nominal update budget. Parameters for two simple processing baselines, grayscale-percentile thresholding and Gaussian-background residuals, are selected on validation images before test evaluation. Conditions and results appear in S4 and S8.\par

\subsection*{3.3 Public-Image Adaptation and Visibility Stratification}

\addcontentsline{toc}{subsection}{3.3 Public-Image Adaptation and Visibility Stratification}

Public adaptation starts from a separate model trained on the laboratory's integrated original annotations. It has the same architecture as the primary internal models, but uses 168 training, 39 validation, and 29 test images, with approximately 30\% of training images confirmed Kikuchi-negative. It is not one of the four primary models in Section 3.2. Public training images follow a predetermined order that interleaves sample groups and visibility categories, yielding nested sets of 20, 40, 60, 80, 100, 120, 140, 160, 180, 200, 220, and 244 images. Each of three runs at each size reloads the same initial segmentation weights, rather than continuing from the preceding adaptation size. The added grade-regression component is randomly initialized, and all segmentation and auxiliary parameters are updated together. One view is processed per update, with a fixed budget of 24,000 updates. Learning rate and weight decay are both \(10^{-5}\), with no learning-rate schedule or early stopping. Every 400 updates, checkpoint selection uses pooled Kikuchi IoU at threshold 0.45 on fixed validation images rated 2–5 by the first author; ties retain the earlier model. The selected checkpoint is evaluated on the fixed test set after each run, without using test metrics for selection. Three-reader medians are used only for final test stratification, while individual ratings supervise training and determine the validation subset. Public training also traverses all D4 views; a fixed update budget means more presentations per image for smaller sets. Controls matching per-image presentations appear in S5, and the full procedure in Algorithm S4.\par

Visibility grades distinguish clear bands from ambiguous cues, rather than assessing film quality. Grade 5 means completely clear, grade 4 fairly clear, grade 3 basically discernible or distinctly Kikuchi-like, grade 2 faint features insufficient for confirmation, and grade 1 no relevant features. A separate −1 label marks known non-targets such as polycrystalline rings, central diffraction rings, or screen edges. It maps to 1 for ordinal statistics but retains a higher false-positive penalty during training. Figure 3 illustrates these judgments using fixed test images selected from the originals and three-reader ratings without consulting masks or model metrics.\par

\begin{center}\begin{minipage}{\linewidth}\centering
\includegraphics[width=\linewidth,height=0.62\textheight,keepaspectratio]{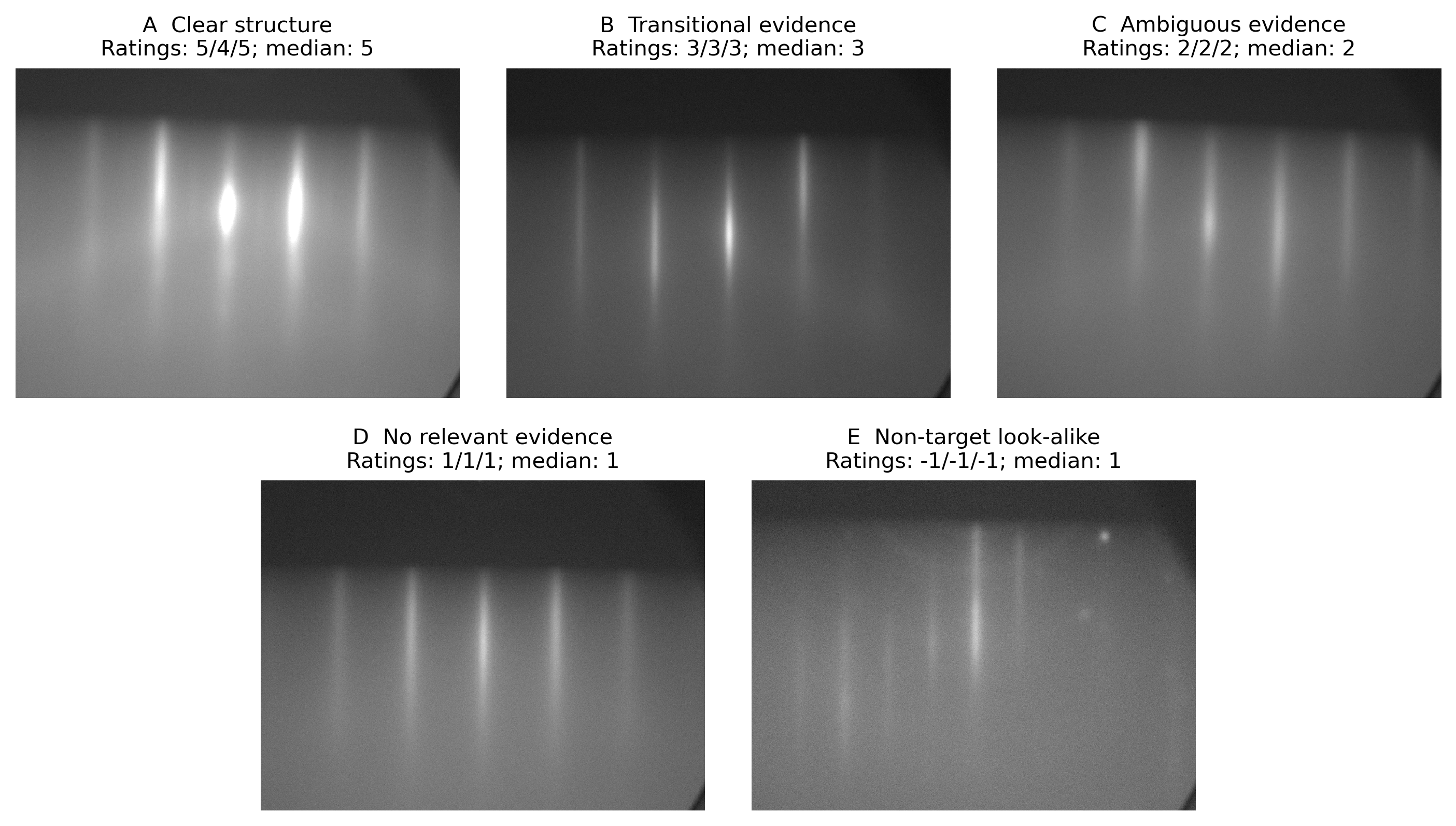}
\par\vspace{3mm}{\small\raggedright Figure 3. Visibility examples from the public test set, displayed with \(\sqrt{x}\) to reveal dark features. A shows clear structures; B discernible or Kikuchi-like structures; C insufficient cues for confirmation; D no relevant features; and E definite non-target structures, such as polycrystalline rings, central diffraction rings, or bright screen edges with other known assignments. The three original ratings shown are from Lumou Weng, Hanshan Huang, and Zhenhan Zhang, in that order. Mapping −1 to 1 before taking the median gives grades 5, 3, 2, 1, and 1, respectively. These examples explain the grades; performance statistics include all test images in each grade. References and predictions for the same five images appear in Figure S11.\par}
\end{minipage}\end{center}

All three raters were graduate students with practical MBE/RHEED experience and judged the images independently. Hanshan Huang and Zhenhan Zhang rated with model outputs hidden, using only the original images and their experience; these two opinions were independent of model predictions and contributed to test stratification. First author Lumou Weng developed the model, knew historical outputs, and also supplied training-image grade supervision. Observers increased display brightness on some images as needed, leaving original data unchanged; per-image brightness values were not recorded during historical grading. Test stratification consistently uses the median of the three mapped grades, combining the two prediction-blinded observers with the developer, retaining ordered integer grades and reducing dependence on one observer's category boundaries. Observer details, dates, and agreement analyses appear in S6.\par

The visibility-assisted adaptation model uses the first author's training-image grades. Grades 5, 4, 3, 2, 1, and −1 assign Kikuchi and union loss weights of 1.00, 0.95, 0.70, 0.40, 1.00, and 2.00, respectively: weak cues participate with reduced weights. These are fixed lookup weights indexed by manual grades and are not dynamically adjusted by predicted grades; the streak loss coefficient is 0.35. The spatial mean of each of the union, streak, and Kikuchi logit maps forms a three-number input to a small learned mapping that outputs an unconstrained continuous auxiliary visibility score, jointly trained against manual grades with a smooth absolute-deviation loss weighted by 0.15. The regression target for −1 is 1, while its segmentation weight remains 2. The continuous output is neither rounded nor clipped to 1–5. It contributes to segmentation learning through parameter updates, but does not gate or rescale masks during inference, and predicted grades do not define test groups. The total objective is\par

\begin{equation*}
\begin{aligned}
L_A={}&w_\ell\bigl(L_{\mathrm{BCE},u}+L_{\mathrm{BCE},k}+0.2L_{D,k}\bigr)\\
&+0.35L_{\mathrm{BCE},s}+0.15L_{\mathrm{reg}}.
\end{aligned}
\end{equation*}

Here \(\ell\) is the first author's manual grade and \(w_\ell\) the corresponding lookup weight. \(L_{\mathrm{reg}}\) is SmoothL1: half the squared error for absolute errors below 1, and absolute error minus 1/2 otherwise. This auxiliary task is part of the fixed adaptation protocol; its independent benefit was not isolated in a separate ablation. Exact structure appears in S2.2 and Algorithms S2 and S4.\par

\subsection*{3.4 Independent Re-annotation and Statistics}

\addcontentsline{toc}{subsection}{3.4 Independent Re-annotation and Statistics}

Fifteen of the 29 internal test images were selected in advance for independent Kikuchi-region annotation by MBE-experienced graduate student Hanshan Huang and supervisor Gan Wang. Their interfaces displayed neither model outputs, historical masks, nor streak labels. Re-annotations were compared with the originals and each of the four fixed models to quantify reference-boundary reproducibility; the original annotations remained the primary references. Forty additional laboratory originals were sampled from batches absent from the training inventories and judged independently by four observers. Ten met the eligibility criterion of at least three negative judgments and no ambiguous or positive judgment, and were used to test extra responses from the fixed models. Selection and width-sensitivity details appear in S4.3 and S7.\par

Means and sample standard deviations across runs describe training randomness. In contrast, 95\% intervals for finite test sources are estimated by resampling with replacement: images or growth batches for internal data and complete sample groups for public data, preserving within-group dependence. Training-size comparisons use paired group draws and paired seeds. These two types of uncertainty are identified separately; random seeds are not additional test samples. Primary statistics evaluate each model before aggregation, while mean-score ensembles used in illustrations are explicitly identified. Resampling procedures and counts appear in S12.6.\par

\section*{4. Results and Discussion}

\addcontentsline{toc}{section}{4. Results and Discussion}

\subsection*{4.1 Explicit Localization in Laboratory Images}

\addcontentsline{toc}{subsection}{4.1 Explicit Localization in Laboratory Images}

The model extracts Kikuchi regions separately from streaks using original manual annotations. On 29 test images separated by complete growth batch, Kikuchi IoU was 0.6290±0.0137 and F1 0.7722±0.0103, with individual IoUs ranging from 0.6133 to 0.6417 (Table 2). Kikuchi precision and recall were 0.7618 and 0.7836, respectively, indicating recovery of substantial reference area with limited extra responses. Streak IoU was 0.6765±0.0114; Figure 1 shows the spatial relationship between the two outputs. Complete precision/recall and run-level statistics appear in Table S4.\par

\par\noindent\begin{minipage}{\linewidth}\small
Table 2. Localization on 29 internal test images against original manual polygons. Values are means±sample standard deviations of pooled pixel-level results from four independent training runs.\par\vspace{4pt}
\setlength{\tabcolsep}{3pt}
\renewcommand{\arraystretch}{1.2}
\begin{tabular}{@{}>{\raggedright\arraybackslash}p{\dimexpr 0.490196\linewidth-2\tabcolsep\relax}>{\raggedright\arraybackslash}p{\dimexpr 0.254902\linewidth-2\tabcolsep\relax}>{\raggedright\arraybackslash}p{\dimexpr 0.254902\linewidth-2\tabcolsep\relax}@{}}
\toprule
\textbf{Target} & \textbf{IoU} & \textbf{F1} \\
\midrule
Candidate Kikuchi regions & 0.6290±0.0137 & 0.7722±0.0103 \\
Streaks & 0.6765±0.0114 & 0.8070±0.0081 \\
\bottomrule
\end{tabular}
\end{minipage}\par

Equally weighted image IoU was 0.6236 (95\% interval 0.5969–0.6511), and equally weighted batch IoU was 0.6183 (0.5900–0.6476), both close to the pooled result. Performance is therefore not dominated by a few large regions or batches with more images. Batch distributions and fixed low-, middle-, and high-scoring examples are retained in the supplement. Evaluating the same models at threshold 0.50 gave IoU 0.6283±0.0128, differing from the 0.45 operating point by 0.0007. The two validation-tuned simple processing baselines gave IoUs of 0.2308 and 0.1437, mainly incorporating bright structures or omitting weak lines. They illustrate the limits of brightness- or residual-based selection on these data, rather than representing all conventional methods.\par

Because the primary internal dataset favors Kikuchi-positive images, we separately examined extra responses on 10 images confirmed negative by multiple observers. False-positive area ranged from 0.394\% to 2.957\% across the four models and was 0.624\% for the mean-score ensemble, mainly around streak ends, central bright spots, and screen edges, with variation across runs (Figure S3). Localization on positive images and response control on negative images therefore warrant separate assessment.\par

\subsection*{4.2 Training-Set Size and Visibility in Public-Image Adaptation}

\addcontentsline{toc}{subsection}{4.2 Training-Set Size and Visibility in Public-Image Adaptation}

The public Streaks images include clear Kikuchi bands, ambiguous cues, and images without targets. We examine overall localization together with strata defined by the three-reader median, distinguishing the evidence in an image before assessing the role of additional training data. Direct application of the laboratory starting model to 86 public test images gave IoU 0.1413. Adaptation with 20, 40, and 60 public training images gave 0.3214±0.0119, 0.3308±0.0131, and 0.3871±0.0079, respectively. From 80 to 244 images, overall point estimates fluctuated locally between 0.4062 and 0.4386 (Figure 4a). The 100-image result was 0.4241±0.0040, compared with 0.4298±0.0290 for the full 244-image set. The paired sample-group 95\% interval for 100 minus 244 images was [−0.0243,\allowbreak{}+0.0172]. Under this fixed partition and update budget, smaller training sets captured much of the overall improvement; subsequent expansion should also be assessed through visibility strata and no-target responses.\par

\begin{center}\begin{minipage}{\linewidth}\centering
\includegraphics[width=\linewidth,height=0.62\textheight,keepaspectratio]{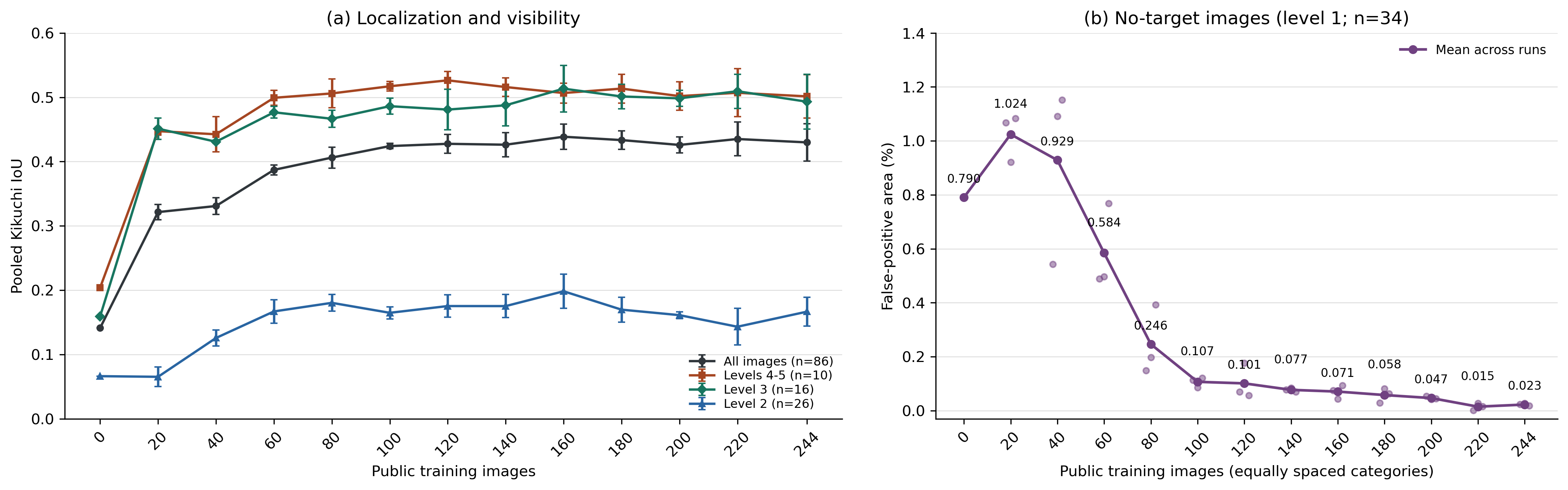}
\par\vspace{3mm}{\small\raggedright Figure 4. Localization and false positives during adaptation to public images. (a) Overall Kikuchi IoU and strata defined by three-reader median grades. Horizontal spacing follows the actual training-image counts; error bars are sample standard deviations across three runs. (b) False-positive area as a fraction of valid image area on 34 grade-1 images with empty references. Training sizes are equally spaced categories labeled with their exact image counts. Large points and lines show means across runs; small points show individual runs. The three small points are offset by 0.1 category spacings around each center solely to prevent overlap and all represent the same training size. Curves include 0, 20, 40, 60, 80, 100, 120, 140, 160, 180, 200, 220, and 244 images. The zero-image point is one laboratory starting model; all adapted sizes share 51 validation images, 86 test images, and 24,000 updates. Complete values and sample-group intervals appear in S5; training processes appear in Figures S4 and S5.\par}
\end{minipage}\end{center}

Visibility better explains differences between patterns than a single overall score. With the full training set, IoU was 0.5014±0.0336 for 10 clear images with median grades 4–5, 0.4931±0.0426 for 16 grade-3 images, and 0.1666±0.0224 for 26 grade-2 images (Table 3). The fixed test set includes 38 prediction-initialized references without confirmed manual review; all 10 clear-group references belong to the earlier manual-annotation cohort (Table S1a). Ten grade-2 images have nonempty Kikuchi references, so their low score is not caused by assigning zero to correctly negative images; weak-line omissions, choices of region extent, and extra predictions all contribute. Images with clear streaks but no Kikuchi features are retained separately to distinguish overall pattern quality from Kikuchi visibility.\par

\par\noindent\begin{minipage}{\linewidth}\small
Table 3. Three-reader median strata with the full 244-image public training set. IoUs are means±sample standard deviations across three runs; no-target images are assessed by false-positive area. The 86 test images come from 12 sample groups; full source composition appears in Table S2.\par\vspace{4pt}
\setlength{\tabcolsep}{3pt}
\renewcommand{\arraystretch}{1.2}
\begin{tabular}{@{}>{\raggedright\arraybackslash}p{\dimexpr 0.493827\linewidth-2\tabcolsep\relax}>{\raggedright\arraybackslash}p{\dimexpr 0.185185\linewidth-2\tabcolsep\relax}>{\raggedright\arraybackslash}p{\dimexpr 0.320988\linewidth-2\tabcolsep\relax}@{}}
\toprule
\textbf{Visibility group} & \textbf{Test images (nonempty/\allowbreak{}empty)} & \textbf{Kikuchi IoU or false-positive area} \\
\midrule
Clear, grades 4–5 & 10 (10/0) & IoU 0.5014±0.0336 \\
Discernible/Kikuchi-like, grade 3 & 16 (14/2) & IoU 0.4931±0.0426 \\
Ambiguous, grade 2 & 26 (10/16) & IoU 0.1666±0.0224 \\
No target, grade 1 (including mapped −1) & 34 (0/34) & False-positive area 0.023\% \\
All images & 86 (34/52) & IoU 0.4298±0.0290 \\
\bottomrule
\end{tabular}
\end{minipage}\par

For clear structures, training with 60, 80, and 100 images yielded IoUs of 0.4992±0.0116, 0.5060±0.0224, and 0.5171±0.0073, respectively, compared with 0.5014±0.0336 for all 244 images. These values provide an empirical reference for additional annotation-set size under the current image source and fixed experimental conditions. No-target responses generally declined as the dataset expanded, with local fluctuations: grade-1 false-positive area was 0.929\% at 40 images, 0.246\% at 80, 0.107\% at 100, and 0.023\% with full training (Figure 4b). Even when localization overlap is similar, a larger training set can improve exclusion of non-target structures.\par

Annotation resources must be counted for the complete experiment. Figure 4 reports training images only; 51 validation and 86 test images are also required. Since the current references include prediction drafts and unreviewed records, the curve describes image requirements under fixed-reference conditions rather than the person-hours for wholly manual annotation. In the control matching presentations per image, extending the 244-image run to 58,560 updates increased IoU by only 0.0098, with a 95\% interval of [−0.0015,\allowbreak{}+0.0235]. Limited training may account for small fluctuations, while the present data do not identify the dominant cause of the plateau. The public results provide a resource reference for supervised adaptation; a prospective external evaluation would require separately established independent references.\par

\subsection*{4.3 Ambiguous Cues, Reference Width, and Methodological Basis}

\addcontentsline{toc}{subsection}{4.3 Ambiguous Cues, Reference Width, and Methodological Basis}

Kikuchi visibility changes continuously, whereas region evaluation requires binary boundaries. Whether a weak segment extends further, whether the inner part of a band is included, and whether obscured segments are connected can vary with viewing conditions and annotation conventions. Although the observers did not always assign identical grades, 93.5\%, 87.7\%, and 85.9\% of their pairwise judgments differed by at most one grade, indicating a shared ordinal trend (S6). The median retains clear, transitional, and ambiguous distinctions, while image-level grades and pixel-boundary consistency describe different aspects of annotation.\par

Independent pixel re-annotation further shows the influence of linewidth conventions. On 15 blinded images, the two independent observers had IoU 0.5516; original annotations agreed with them at 0.4026 and 0.4102, while model-to-observer IoU ranged from 0.3911 to 0.4154. The independent observers' regions covered approximately twice the area of the originals and model outputs. Narrower labels and predictions largely lay within the broader regions, yielding high precision but low recall. The model therefore primarily reproduces the narrower cores defined by the training references, rather than the full visible widths chosen by the independent observers. Shrinking the wider references inward while fixing predictions only partly improved overlap (S7); both reference width and genuine localization errors contribute to the differences.\par

Technically, the workflow first ensures that weak Kikuchi regions enter supervision and that both positive requirements are retained at intersections. Multiscale responses and skip-connected detail provide different spatial scopes for localization. The strong-background/weak-target problem inspired by LarkMNet is implemented through image representation, network paths, and full-image region supervision. In six matched controls, D4 and soft-Dice corresponded to mean pooled IoU increments of 0.0091 and 0.0056, respectively; the absolute differences for other changes were at most 0.0057, and most batch-resampling intervals included zero (S8). The complete workflow achieves Kikuchi localization, while these tests do not consistently attribute the gain to a single added module. In particular, they do not establish the attention branch as the decisive source.\par

\subsection*{4.4 Uses and Conditions for RHEED Measurement}

\addcontentsline{toc}{subsection}{4.4 Uses and Conditions for RHEED Measurement}

Explicit region outputs allow subsequent analysis to focus on specific lines and bands. Original grayscale profiles can be sampled perpendicular to features near the predictions to fit centers and edges. Separate streak and Kikuchi regions also identify intersection windows requiring inspection. Previous studies observed Kikuchi-related intensity enhancement at streak intersections and effects on intensity oscillations [7,8], giving these overlapping positions measurement relevance. Region area defines localization windows; separating intensity contributions requires the original grayscale data and analysis of scattering processes.\par

Obtaining plane directions, mean inner potential, or lattice deformation from positions requires calibration from screen coordinates to diffraction angles, sample orientation, and an RHEED-appropriate description of refraction and surface resonance [3–6,20,21]. The present work establishes inspectable region localization. A next step is to formulate a specific physical measurement around changes in line positions, band shapes, or intersections within degassing, growth, and annealing sequences from the same batch. Real-time use further requires assessment of acquisition latency, long-term brightness drift, and localization-window stability. Current results apply to static datasets dominated by streak patterns; extension to polycrystalline, spot-like, or abnormal patterns remains to be evaluated.\par

\section*{5. Conclusions}

\addcontentsline{toc}{section}{5. Conclusions}

We developed a workflow that localizes streak and Kikuchi regions separately while allowing overlap, achieving Kikuchi IoU 0.6290±0.0137 on laboratory images with original manual references and growth-batch-separated partitions. Supervised adaptation to public chalcogenide RHEED images provides an empirical reference for additional image requirements: 100 training images gave clear-group IoU 0.5171±0.0073, while more data reduced no-target responses. Agreement was lower for ambiguous cues. Three-reader visibility stratification and independent region re-annotation reveal distinct effects of perceptual visibility and linewidth conventions on evaluation; public values refer specifically to this study's mixed-provenance fixed references. The workflow provides a practical route to extracting weak Kikuchi spatial structures from bright streak backgrounds and an inspectable starting point for subsequent analysis of line shapes, intersections, and local intensity.\par

\section*{Acknowledgments and Author Declarations}

\addcontentsline{toc}{section}{Acknowledgments and Author Declarations}

We thank Wenyu Hu, Jie Jiang, and Zijun Yan for their MBE experiments and accumulation of RHEED images, which contributed to the laboratory dataset. We thank Hangyu Yin for independent image assessment, and N. Muetzel and colleagues for releasing chalcogenide RHEED images and code [18,19], which enabled adaptation to a different image source. This work received no dedicated external funding and used the group's existing experimental and computational resources. The authors declare no competing interests.\par

Author contributions (CRediT): Lumou Weng: Conceptualization, Methodology, Software, Investigation (including RHEED image acquisition), Data curation, Validation, Formal analysis, Visualization, Writing – original draft, and Writing – review and editing. Hanshan Huang: Investigation (RHEED image acquisition, independent image assessment, and region re-annotation), Validation, and Data curation. Zhenhan Zhang: Validation (independent image assessment) and Data curation (visibility ratings). Gan Wang: Supervision, Resources, Validation, and Writing – review and editing. Multi-observer assessment concerned material-characterization images containing no personal information and involved no human specimens or animal experiments.\par

\section*{Data, Code, and AI-Assistance Statement}

\addcontentsline{toc}{section}{Data, Code, and AI-Assistance Statement}

Public source images and their original classification code are available in Ref. [19] under the Creative Commons Attribution 4.0 International license (CC BY 4.0). Public-source RHEED images reproduced here are credited to that dataset; display transformations and reference/prediction overlays are described in the figure captions. The present study's region annotations, visibility records, fixed partitions, model weights, and training and analysis scripts are retained in the local project archive and are not publicly distributed with this preprint. Supplementary Information S12 provides pseudocode for preprocessing, model computation, training, and evaluation.\par

Generative AI tools, including OpenAI ChatGPT/Codex, Claude, and Grok, assisted literature organization, software development and review, statistical plotting scripts, and manuscript revision. Training, inference, and numerical statistics were executed on local data, and figures use real RHEED images and computed model outputs. The authors are responsible for verifying scientific content and sources and approving the final manuscript; AI tools are not authors. Tool uses and version records are retained in the project archive.\par

\section*{References}

\addcontentsline{toc}{section}{References}

\par\begingroup\interlinepenalty=10000\noindent\hangindent=2em\hangafter=1 [1] S. Miyake, K. Hayakawa, T. Kawamura, Y.-H. Ohtsuki. The Nature of Kikuchi Lines in the Bragg Case in Reflexion High-Energy Electron Diffraction. Acta Crystallographica A 31, 32-38 (1975). \href{https://doi.org/10.1107/S0567739475000058}{DOI}.\par\endgroup

\par\begingroup\interlinepenalty=10000\noindent\hangindent=2em\hangafter=1 [2] J. Pawlak, M. Przybylski, Z. Mitura. An Analysis of Kikuchi Lines Observed with a RHEED Apparatus for a TiO2-Terminated SrTiO3 (001) Crystal. Materials 14, 7077 (2021). \href{https://doi.org/10.3390/ma14227077}{DOI}.\par\endgroup

\par\begingroup\interlinepenalty=10000\noindent\hangindent=2em\hangafter=1 [3] K. Hayakawa, N. Aizawa. Identification of Crystal Axes by Kikuchi Patterns in Reflection High-Energy Electron Diffraction. Japanese Journal of Applied Physics 21, 1105 (1982). \href{https://doi.org/10.1143/JJAP.21.1105}{DOI}.\par\endgroup

\par\begingroup\interlinepenalty=10000\noindent\hangindent=2em\hangafter=1 [4] T. Gotoh, S. Murakami, K. Kinosita, Y. Murata. Surface Rumpling of MgO(001) Deduced from Changes in RHEED Kikuchi Pattern. I. Experimental. Journal of the Physical Society of Japan 50, 2063-2068 (1981). \href{https://doi.org/10.1143/JPSJ.50.2063}{DOI}.\par\endgroup

\par\begingroup\interlinepenalty=10000\noindent\hangindent=2em\hangafter=1 [5] Y. Murata. Surface Rumpling of MgO(001) Deduced from Change in RHEED Kikuchi Pattern. II. Theoretical. Journal of the Physical Society of Japan 51, 1932-1935 (1982). \href{https://doi.org/10.1143/JPSJ.51.1932}{DOI}.\par\endgroup

\par\begingroup\interlinepenalty=10000\noindent\hangindent=2em\hangafter=1 [6] M. Inoue, K. Ueda. Interaction between Kikuchi Lines and Surface Waves in RHEED Pattern of Si(001) Surface. Japanese Journal of Applied Physics 25, 165-166 (1986). \href{https://doi.org/10.1143/JJAP.25.165}{DOI}.\par\endgroup

\par\begingroup\interlinepenalty=10000\noindent\hangindent=2em\hangafter=1 [7] P. K. Larsen et al. Surface disorder induced Kikuchi features in reflection high-energy electron diffraction patterns of static and growing GaAs(001) films. Journal of Vacuum Science \& Technology A 5, 611-614 (1987). \href{https://doi.org/10.1116/1.574684}{DOI}.\par\endgroup

\par\begingroup\interlinepenalty=10000\noindent\hangindent=2em\hangafter=1 [8] G. E. Crook et al. Effects of Kikuchi scattering on reflection high-energy electron diffraction intensities during molecular-beam epitaxy GaAs growth. Journal of Vacuum Science \& Technology A 7, 2549-2553 (1989). \href{https://doi.org/10.1116/1.575795}{DOI}.\par\endgroup

\par\begingroup\interlinepenalty=10000\noindent\hangindent=2em\hangafter=1 [9] A. J. Wilkinson, T. B. Britton. Strains, planes, and EBSD in materials science. Materials Today 15, 366-376 (2012). \href{https://doi.org/10.1016/S1369-7021\%2812\%2970163-3}{DOI}.\par\endgroup

\par\begingroup\interlinepenalty=10000\noindent\hangindent=2em\hangafter=1 [10] S. I. Wright, M. M. Nowell, D. P. Field. A Review of Strain Analysis Using Electron Backscatter Diffraction. Microscopy and Microanalysis 17, 316-329 (2011). \href{https://doi.org/10.1017/S1431927611000055}{DOI}.\par\endgroup

\par\begingroup\interlinepenalty=10000\noindent\hangindent=2em\hangafter=1 [11] A. Winkelmann et al. Kikuchi pattern simulations of backscattered and transmitted electrons. Journal of Microscopy 284, 157-184 (2021). \href{https://doi.org/10.1111/jmi.13051}{DOI}.\par\endgroup

\par\begingroup\interlinepenalty=10000\noindent\hangindent=2em\hangafter=1 [12] R. K. Vasudevan et al. Big-Data Reflection High Energy Electron Diffraction Analysis for Understanding Epitaxial Film Growth Processes. ACS Nano 8, 10899-10908 (2014). \href{https://doi.org/10.1021/nn504730n}{DOI}.\par\endgroup

\par\begingroup\interlinepenalty=10000\noindent\hangindent=2em\hangafter=1 [13] S. R. Provence et al. Machine Learning Analysis of Perovskite Oxides Grown by Molecular Beam Epitaxy. Physical Review Materials 4, 083807 (2020). \href{https://doi.org/10.1103/PhysRevMaterials.4.083807}{DOI}.\par\endgroup

\par\begingroup\interlinepenalty=10000\noindent\hangindent=2em\hangafter=1 [14] H. Liang et al. Application of machine learning to reflection high-energy electron diffraction images for automated structural phase mapping. Physical Review Materials 6, 063805 (2022). \href{https://doi.org/10.1103/PhysRevMaterials.6.063805}{DOI}.\par\endgroup

\par\begingroup\interlinepenalty=10000\noindent\hangindent=2em\hangafter=1 [15] C. C. Price et al. Predicting and Accelerating Nanomaterial Synthesis Using Machine Learning Featurization. Nano Letters 24, 14862-14867 (2024). \href{https://doi.org/10.1021/acs.nanolett.4c04500}{DOI}.\par\endgroup

\par\begingroup\interlinepenalty=10000\noindent\hangindent=2em\hangafter=1 [16] A. Khaireh-Walieh et al. Data-Driven Azimuthal RHEED Construction for In Situ Crystal Growth Characterization. Crystal Growth \& Design 25, 7438-7445 (2025). \href{https://doi.org/10.1021/acs.cgd.5c00368}{DOI}.\par\endgroup

\par\begingroup\interlinepenalty=10000\noindent\hangindent=2em\hangafter=1 [17] P. T. Gemperline, R. Paudel, R. K. Vasudevan, R. B. Comes. Improvement of data analytics techniques in reflection high-energy electron diffraction to enable machine learning. Journal of Vacuum Science \& Technology A 43, 032701 (2025). \href{https://doi.org/10.1116/6.0004400}{DOI}.\par\endgroup

\par\begingroup\interlinepenalty=10000\noindent\hangindent=2em\hangafter=1 [18] N. Muetzel et al. RHEED pattern classification by a convolutional neural network for the growth of chalcogenide thin films and nanostructures. AIP Advances 16, 035317 (2026). \href{https://doi.org/10.1063/5.0308043}{DOI}.\par\endgroup

\par\begingroup\interlinepenalty=10000\noindent\hangindent=2em\hangafter=1 [19] M. Gebran, N. Muetzel, X. Liu, B. Assaf. Dataset and code for [dataset associated with Ref. [18]]. Zenodo (2025), record 17476777, CC BY 4.0. \href{https://doi.org/10.5281/zenodo.17476777}{Data DOI}.\par\endgroup

\par\begingroup\interlinepenalty=10000\noindent\hangindent=2em\hangafter=1 [20] T. B. Britton, A. J. Wilkinson. High resolution electron backscatter diffraction measurements of elastic strain variations in the presence of larger lattice rotations. Ultramicroscopy 114, 82-95 (2012). \href{https://doi.org/10.1016/j.ultramic.2012.01.004}{DOI}.\par\endgroup

\par\begingroup\interlinepenalty=10000\noindent\hangindent=2em\hangafter=1 [21] A. Winkelmann. Dynamical Simulation of Electron Backscatter Diffraction Patterns. In Electron Backscatter Diffraction in Materials Science, Second Edition, pp. 21-33 (2009). \href{https://doi.org/10.1007/978-0-387-88136-2_2}{DOI}.\par\endgroup

\par\begingroup\interlinepenalty=10000\noindent\hangindent=2em\hangafter=1 [22] O. Ronneberger, P. Fischer, T. Brox. U-Net: Convolutional Networks for Biomedical Image Segmentation. In Medical Image Computing and Computer-Assisted Intervention – MICCAI 2015, pp. 234–241 (2015). \href{https://doi.org/10.1007/978-3-319-24574-4_28}{DOI}.\par\endgroup

\par\begingroup\interlinepenalty=10000\noindent\hangindent=2em\hangafter=1 [23] T. C. Kaspar et al. Machine-learning-enabled on-the-fly analysis of RHEED patterns during thin film deposition by molecular beam epitaxy. Journal of Vacuum Science \& Technology A 43, 032702 (2025). \href{https://doi.org/10.1116/6.0004493}{DOI}.\par\endgroup

\par\begingroup\interlinepenalty=10000\noindent\hangindent=2em\hangafter=1 [24] H. Qin et al. Superconductivity in Single-Quintuple-Layer Bi2Te3 Grown on Epitaxial FeTe. Nano Letters 20, 3160-3168 (2020). \href{https://doi.org/10.1021/acs.nanolett.9b05167}{DOI}.\par\endgroup

\par\begingroup\interlinepenalty=10000\noindent\hangindent=2em\hangafter=1 [25] J. Chen et al. Evidence for Magnetic Skyrmions at the Interface of Ferromagnet/Topological-Insulator Heterostructures. Nano Letters 19, 6144-6151 (2019). \href{https://doi.org/10.1021/acs.nanolett.9b02191}{DOI}.\par\endgroup

\par\begingroup\interlinepenalty=10000\noindent\hangindent=2em\hangafter=1 [26] W. Hu et al. Interfacial Reaction-Associated Sample Quality and Superconductivity in Bi2Te3–FeTe Thin Films. ACS Applied Electronic Materials 7, 4382-4390 (2025). \href{https://doi.org/10.1021/acsaelm.5c00588}{DOI}.\par\endgroup

\par\begingroup\interlinepenalty=10000\noindent\hangindent=2em\hangafter=1 [27] H. Li et al. Molecular Beam Epitaxy Grown Cr2Te3 Thin Films with Tunable Curie Temperatures for Spintronic Devices. ACS Applied Nano Materials 2, 6809-6817 (2019). \href{https://doi.org/10.1021/acsanm.9b01179}{DOI}.\par\endgroup

\par\begingroup\interlinepenalty=10000\noindent\hangindent=2em\hangafter=1 [28] J. Chen et al. Conformal Growth of Cr2Te3 on Bi2Te3 Nanodots with a Topological Hall Effect. Crystal Growth \& Design 22, 140-147 (2022). \href{https://doi.org/10.1021/acs.cgd.1c00657}{DOI}.\par\endgroup

\par\begingroup\interlinepenalty=10000\noindent\hangindent=2em\hangafter=1 [29] J. Li, X. Lu, P. Zhang, Q. Li. Weak feature crack detection in high-resolution concrete dam surface image with LarkMNet. Measurement 238, 115327 (2024). \href{https://doi.org/10.1016/j.measurement.2024.115327}{DOI}.\par\endgroup

\par\begingroup\interlinepenalty=10000\noindent\hangindent=2em\hangafter=1 [30] Z. Mitura, G. Szwachta, Ł. Kokosza, M. Przybylski. Identification of Kikuchi lines in electron diffraction patterns collected in small-angle geometry. Acta Crystallographica A 80, 104–111 (2024). \href{https://doi.org/10.1107/S2053273323009385}{DOI}.\par\endgroup

\clearpage

\section*{Supplementary Information}

\addcontentsline{toc}{section}{Supplementary Information}

The supplement uses the same data and reference versions as the main text and provides complete settings, statistics, and diagnostics. Image-level grading and region annotation are separate tasks; development validation, internal testing against original annotations, and public adaptation are reported separately. Supplementary figure and table numbering is independent of the main text and uses the prefix S.\par

\subsection*{S1. Images, Reference Provenance, and Preprocessing}

\addcontentsline{toc}{subsection}{S1. Images, Reference Provenance, and Preprocessing}

\subsubsection*{S1.1 Internal Annotations and the Public-Adaptation Starting Model}

\addcontentsline{toc}{subsubsection}{S1.1 Internal Annotations and the Public-Adaptation Starting Model}

The original polygons in Section 3.1 were rasterized into separate binary streak and Kikuchi masks. Regions of the same class were combined by union, and intersections between classes were retained in both masks. The same version of all 290 annotation files was used throughout primary training, validation, and evaluation.\par

The integrated inventory used for the public-adaptation starting model contains 354 images; actual partitions and confirmed-negative proportions are given in Section 3.3. Early images with streak annotations but unknown Kikuchi status were listed separately, rather than automatically treated as negatives. The earlier six-class annotations belong to method development; six mask files do not represent six independent images.\par

\subsubsection*{S1.2 Public Reference Provenance and Counts}

\addcontentsline{toc}{subsubsection}{S1.2 Public Reference Provenance and Counts}

Public-image selection and group separation are described in Section 3.1. Tables S1 and S2 list the provenance of the fixed region references and test-set composition.\par

\par\noindent\begin{minipage}{\linewidth}\small
Table S1. Region-reference provenance for eligible public images. Status follows the fixed records for this analysis, rather than being inferred from directory names.\par\vspace{4pt}
\setlength{\tabcolsep}{3pt}
\renewcommand{\arraystretch}{1.2}
\begin{tabular}{@{}>{\raggedright\arraybackslash}p{\dimexpr 0.438095\linewidth-2\tabcolsep\relax}>{\raggedright\arraybackslash}p{\dimexpr 0.123810\linewidth-2\tabcolsep\relax}>{\raggedright\arraybackslash}p{\dimexpr 0.438095\linewidth-2\tabcolsep\relax}@{}}
\toprule
\textbf{Provenance} & \textbf{Count} & \textbf{Manual review status} \\
\midrule
Earlier manual region annotations & 203 & Existing manual annotations \\
Prediction drafts with subsequent manual review & 40 & Reviewed \\
Prediction-initialized drafts & 138 & Not yet manually reviewed in the fixed records \\
Total & 381 & Fixed references of mixed provenance \\
\bottomrule
\end{tabular}
\end{minipage}\par

Provenance was checked against fixed records, not directory names. Three-reader image-level grading and pixel-reference construction are separate tasks; the scope of the public statistics is stated in Section 3.1. Table S1a connects reference provenance to the visibility strata of the fixed test set.\par

\par\noindent\begin{minipage}{\linewidth}\small
Table S1a. Review status of the fixed public test references by three-reader median grade. The first column of counts combines earlier manual annotations and manually reviewed prediction drafts. Review status follows the fixed provenance records.\par\vspace{4pt}
\setlength{\tabcolsep}{3pt}
\renewcommand{\arraystretch}{1.2}
\begin{tabular}{@{}>{\raggedright\arraybackslash}p{\dimexpr 0.337349\linewidth-2\tabcolsep\relax}>{\raggedright\arraybackslash}p{\dimexpr 0.289157\linewidth-2\tabcolsep\relax}>{\raggedright\arraybackslash}p{\dimexpr 0.216867\linewidth-2\tabcolsep\relax}>{\raggedright\arraybackslash}p{\dimexpr 0.156627\linewidth-2\tabcolsep\relax}@{}}
\toprule
\textbf{Median-grade group} & \textbf{Existing manual/completed-review records} & \textbf{Prediction drafts without confirmed review} & \textbf{Total} \\
\midrule
Clear, grades 4–5 & 10 & 0 & 10 \\
Grade 3 & 12 & 4 & 16 \\
Grade 2 & 11 & 15 & 26 \\
Grade 1, including mapped −1 & 15 & 19 & 34 \\
All test images & 48 & 38 & 86 \\
\bottomrule
\end{tabular}
\end{minipage}\par

Of the 48 test references with existing manual or completed-review records, 40 belong to the earlier 203-image annotation cohort and eight to the 40-image reviewed-draft cohort. All 10 clear-group references belong to the former. Earlier annotations also include model-assisted editing during development; the provenance categories describe the fixed records and do not establish prediction-independent pixel ground truth.\par

\par\noindent\begin{minipage}{\linewidth}\small
Table S2. Median-grade composition of the fixed 86-image public test set. The −1 label is mapped to 1 before taking the median of the three ratings.\par\vspace{4pt}
\setlength{\tabcolsep}{3pt}
\renewcommand{\arraystretch}{1.2}
\begin{tabular}{@{}>{\raggedright\arraybackslash}p{\dimexpr 0.173333\linewidth-2\tabcolsep\relax}>{\raggedright\arraybackslash}p{\dimexpr 0.173333\linewidth-2\tabcolsep\relax}>{\raggedright\arraybackslash}p{\dimexpr 0.240000\linewidth-2\tabcolsep\relax}>{\raggedright\arraybackslash}p{\dimexpr 0.240000\linewidth-2\tabcolsep\relax}>{\raggedright\arraybackslash}p{\dimexpr 0.173333\linewidth-2\tabcolsep\relax}@{}}
\toprule
\textbf{Median grade} & \textbf{Images} & \textbf{Nonempty Kikuchi references} & \textbf{Empty references} & \textbf{Sample groups} \\
\midrule
4–5 & 10 & 10 & 0 & 7 \\
3 & 16 & 14 & 2 & 8 \\
2 & 26 & 10 & 16 & 6 \\
1 & 34 & 0 & 34 & 5 \\
All & 86 & 34 & 52 & 12 \\
\bottomrule
\end{tabular}
\end{minipage}\par

Images in the same group may have different visibility, so a group may appear in several rows; there are 12 distinct groups overall. Metrics follow Section 2.4.\par

\subsubsection*{S1.3 Numerical Reading, Interpolation, and Display}

\addcontentsline{toc}{subsubsection}{S1.3 Numerical Reading, Interpolation, and Display}

Internal training and per-image audits read images in OpenCV grayscale mode, obtaining 8-bit grayscale values that are divided by 255. The 512×512 images are resized to 256×256 using default bilinear interpolation; masks use nearest-neighbor interpolation. The original internal test entry used area interpolation for grayscale resizing. Checking all 145 originals confirmed that, for exact halving from 512×512 to 256×256, both methods produce pixel-wise identical 8-bit arrays. The public program reads with the original data type preserved and converts color images to grayscale. Eight-bit values are divided by 255 and 16-bit values by 65535, then clipped to 0–1. This is the actual reading protocol and does not imply that all internal archives are 16-bit.\par

Public resizing uses \(r=\min(256/W,256/H)\), with resized dimensions rounded to integers, centered zero padding to 256×256, and the four padding amounts recorded. Both grayscale images and binary masks use area interpolation; resized masks are binarized at 0.5. D4 applies identical transformations to the image and both masks. Predictions are first thresholded at 0.45 on the 256 grid, then unpadded and restored to original dimensions by nearest-neighbor interpolation. Statistics use the valid original-image grid; internal statistics use the 256 grid directly.\par

Figures 1 and 3 transform display brightness \(x\) to \(\sqrt{x}\). Since \(\sqrt{x}>x\) for \(0<x<1\), darker features are enhanced more strongly. This changes neither image files, model inputs, nor evaluation. Figure S11 uses a separately specified percentile stretch, also solely for readability. Except for explicitly separated development controls, display brightening is not a training augmentation.\par

\subsection*{S2. Network, Training Objectives, and Method Development}

\addcontentsline{toc}{subsection}{S2. Network, Training Objectives, and Method Development}

\subsubsection*{S2.1 Exact Computation of the Compressed Context Branch}

\addcontentsline{toc}{subsubsection}{S2.1 Exact Computation of the Compressed Context Branch}

The base model follows Figure 2, with two main-path max-pooling steps and 48/\allowbreak{}96/\allowbreak{}192 channels. The two reconstruction concatenations have 192+96 and 96+48 input channels. The 3×3 convolutions use stride 1 and padding 1; upsampling is bilinear with corner alignment disabled. Final 1×1 convolutions map 48→1 for the union head and 48→2 for the class head. The classes have independent sigmoid outputs, and the union threshold does not gate Kikuchi predictions.\par

The bottleneck \(B\in\mathbb R^{64\times64\times192}\) passes through average pooling with window and stride 8 and a 1×1 convolution to produce 8×8×64. Its 64 positions are reshaped into 64-dimensional vectors and passed through LayerNorm, forming matrix \(Z\), whose rows represent positions and columns feature components. Single-head attention is\par

\begin{equation*}
Q=ZW_Q,\quad K=ZW_K,\quad V=ZW_V,
\end{equation*}

\begin{equation*}
A=\operatorname{softmax}_{\mathrm{row}}(QK^T/\sqrt{64}),\qquad Y=AVW_O.
\end{equation*}

Bias terms omitted from these equations are enabled in the implementation. Softmax normalizes each row; \(A\) describes weights between positions, rather than between channels. The output is reshaped to 8×8×64, mapped back to 192 channels by a 1×1 convolution, and upsampled eightfold:\par

\begin{equation*}
\widetilde B=B+\gamma\operatorname{Up}_8[\operatorname{Conv}_{1\times1}(\operatorname{reshape}Y)].
\end{equation*}

The trainable scalar \(\gamma\) is initialized to 0. Dropout2d with probability 0.05 is applied after this branch and disabled during inference. The branch has 41,601 parameters, approximately 3.78\% of the total. Compression reduces the positional association matrix from 4096×4096 to 64×64, allowing low-cost exchange across image regions.\par

\subsubsection*{S2.2 Auxiliary Visibility Objective for Public Images}

\addcontentsline{toc}{subsubsection}{S2.2 Auxiliary Visibility Objective for Public Images}

The public model retains the base segmentation network. The spatial means of its union, streak, and Kikuchi logit maps form a three-dimensional input to a 3→24 fully connected layer, ReLU, and a 24→1 fully connected layer that predicts a continuous grade. This output is an auxiliary ordinal description, not a calibrated physical intensity. Manual grades 1–5 specify order without assuming equally spaced physical quantities.\par

The auxiliary head adds 121 parameters. Grade mappings, region weights, and the SmoothL1 objective are given in Section 3.3; implementation appears in Algorithms S2 and S4.\par

\subsubsection*{S2.3 LarkMNet Inspiration and Supervision Checks}

\addcontentsline{toc}{subsubsection}{S2.3 LarkMNet Inspiration and Supervision Checks}

LarkMNet [29] localizes weak dam-surface cracks with detection boxes, using a large-kernel LarkNet, multilevel aggregation and distribution, a shared detection head, and box-regression losses. We draw on its principle of retaining local and contextual cues for weak targets against strong backgrounds, implementing pixel-wise localization through U-Net-style encoding and reconstruction, a compressed-context branch, and independent two-class supervision. The respective outputs are detection boxes and pixel regions, with different implementation and evaluation tasks.\par

Early code included only the streak class in the loss. Kikuchi supervision was subsequently added explicitly, followed by a change from mutually exclusive class supervision to two independent BCE objectives. In a synthetic gradient check, an overlapping reference pixel was \((1,1)\) with both logits 0. The early argmax-based exclusive target gave gradients \((-0.5,+0.5)\), increasing streak response while decreasing Kikuchi response. Independent BCE gave \((-0.25,-0.25)\), increasing both. This check verifies supervision logic rather than measuring a performance increment.\par

Development records before soft-Dice used 116 original images expanded to 928 D4 views, with 29 images serving for validation and checkpoint selection. The lowest validation loss, 0.2214 at epoch 35, corresponded to Kikuchi IoU 0.5951; the highest recorded Kikuchi IoU was 0.6103 at epoch 54, at threshold 0.50. These are development-validation scores, distinct from the batch-separated test results in the main text. A historical note giving a value near 0.64 referred to the union and is likewise not a Kikuchi test metric.\par

\subsection*{S3. Internal Training Process and Computational Resources}

\addcontentsline{toc}{subsection}{S3. Internal Training Process and Computational Resources}

Primary internal training uses batch size 1, AdamW, initial learning rate \(8\times10^{-5}\), and weight decay \(10^{-4}\). A validation-loss plateau with patience 8 halves the learning rate, down to \(2\times10^{-5}\). Checkpoints are selected by validation Kikuchi F1 at threshold 0.50, with early stopping after 25 consecutive epochs without a new best value and a maximum of 80 epochs. Mixed precision uses lower numerical precision for some operations to reduce GPU memory use, without changing the objective definition.\par

\par\noindent\begin{minipage}{\linewidth}\small
Table S3. Records of the four internal training runs.\par\vspace{4pt}
\setlength{\tabcolsep}{3pt}
\renewcommand{\arraystretch}{1.2}
\begin{tabular}{@{}>{\raggedright\arraybackslash}p{\dimexpr 0.152941\linewidth-2\tabcolsep\relax}>{\raggedright\arraybackslash}p{\dimexpr 0.200000\linewidth-2\tabcolsep\relax}>{\raggedright\arraybackslash}p{\dimexpr 0.152941\linewidth-2\tabcolsep\relax}>{\raggedright\arraybackslash}p{\dimexpr 0.282353\linewidth-2\tabcolsep\relax}>{\raggedright\arraybackslash}p{\dimexpr 0.211765\linewidth-2\tabcolsep\relax}@{}}
\toprule
\textbf{Seed} & \textbf{Epochs completed} & \textbf{Saved epoch} & \textbf{Learning-rate reduction epochs} & \textbf{Time between first- and last-epoch records} \\
\midrule
42 & 80 & 67 & 63, 72 & 39 min 11 s \\
7 & 75 & 50 & 45, 59 & 42 min 45 s \\
123 & 70 & 45 & 33, 54 & 33 min 08 s \\
2026 & 80 & 56 & 36, 58 & 29 min 57 s \\
\bottomrule
\end{tabular}
\end{minipage}\par

\begin{center}\begin{minipage}{\linewidth}\centering
\includegraphics[width=\linewidth,height=0.62\textheight,keepaspectratio]{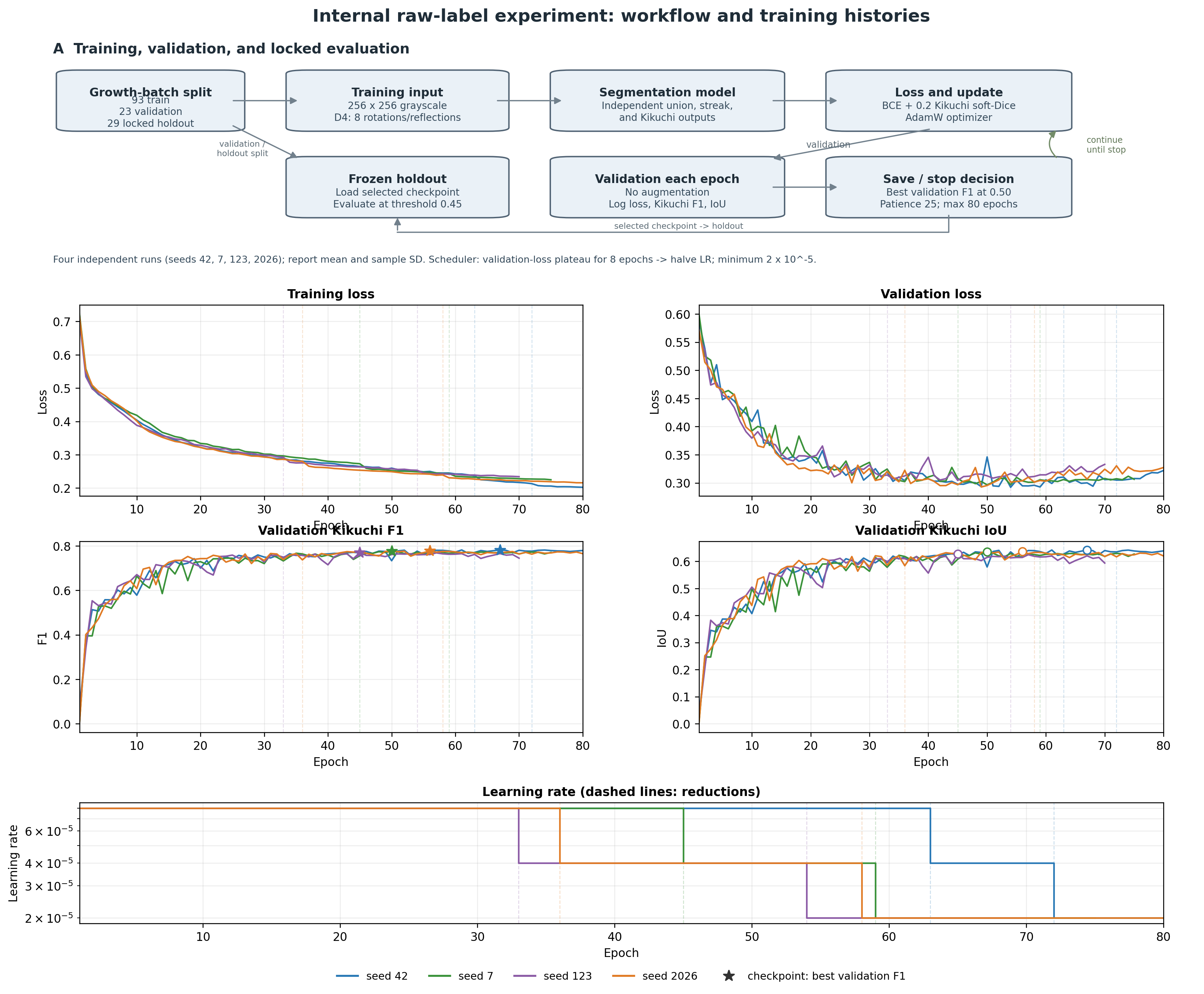}
\par\vspace{3mm}{\small\raggedright Figure S1. Internal training process: batch partitioning, D4 transformation of training images, loss-based updates, epoch-level validation, and checkpoint selection by validation F1, followed by testing at the fixed 0.45 threshold. Curves show training/validation loss, validation Kikuchi F1 and IoU, and learning rate. Stars mark saved epochs and vertical dashed lines learning-rate changes. The figure reads existing logs directly; test scores do not enter the curves or checkpoint selection.\par}
\end{minipage}\end{center}

The computational platform is an RTX 4060 with Python 3.10.19, PyTorch 2.5.1+cu121, and CUDA 12.1. For one 256×256 grayscale image, the fixed model takes a median 7.420 ms from reading and preprocessing through GPU forward calculation and thresholding to a CPU mask, with peak allocated GPU memory 164.8 MiB. This excludes the camera, interface, batch file writing, and manual operations; it measures local computational overhead rather than complete MBE-system response. Historical public runs did not consistently retain complete wall-clock timing, so training times are not ranked.\par

\subsection*{S4. Complete Internal Metrics, Batch Distributions, and Negative Controls}

\addcontentsline{toc}{subsection}{S4. Complete Internal Metrics, Batch Distributions, and Negative Controls}

\subsubsection*{S4.1 Complete Region Statistics}

\addcontentsline{toc}{subsubsection}{S4.1 Complete Region Statistics}

\par\noindent\begin{minipage}{\linewidth}\small
Table S4. Pooled pixel-level results on 29 internal test images against original manual references. The overlap row compares the intersection of the two reference masks with the intersection of the two predicted masks. It describes spatial agreement where both region outputs are present, rather than a scattering-intensity component.\par\vspace{4pt}
\setlength{\tabcolsep}{3pt}
\renewcommand{\arraystretch}{1.2}
\begin{tabular}{@{}>{\raggedright\arraybackslash}p{\dimexpr 0.243243\linewidth-2\tabcolsep\relax}>{\raggedright\arraybackslash}p{\dimexpr 0.229730\linewidth-2\tabcolsep\relax}>{\raggedright\arraybackslash}p{\dimexpr 0.175676\linewidth-2\tabcolsep\relax}>{\raggedright\arraybackslash}p{\dimexpr 0.175676\linewidth-2\tabcolsep\relax}>{\raggedright\arraybackslash}p{\dimexpr 0.175676\linewidth-2\tabcolsep\relax}@{}}
\toprule
\textbf{Target/seed} & \textbf{Precision} & \textbf{Recall} & \textbf{F1} & \textbf{IoU} \\
\midrule
Kikuchi/42 & 0.7816 & 0.7819 & 0.7818 & 0.6417 \\
Kikuchi/7 & 0.7303 & 0.8072 & 0.7668 & 0.6218 \\
Kikuchi/123 & 0.7497 & 0.7711 & 0.7603 & 0.6133 \\
Kikuchi/2026 & 0.7857 & 0.7739 & 0.7798 & 0.6390 \\
Kikuchi/mean±SD & 0.7618±0.0265 & 0.7836±0.0164 & 0.7722±0.0103 & 0.6290±0.0137 \\
Streak/mean±SD & 0.8442±0.0348 & 0.7746±0.0295 & 0.8070±0.0081 & 0.6765±0.0114 \\
Overlap/mean±SD & 0.6109±0.0417 & 0.6301±0.0319 & 0.6190±0.0163 & 0.4484±0.0171 \\
\bottomrule
\end{tabular}
\end{minipage}\par

First averaging each image's metrics across the four runs, then weighting images or batches equally, gives image-weighted IoU 0.6236/F1 0.7651 and batch-weighted IoU 0.6183/F1 0.7613. Image- and batch-resampling IoU intervals are [0.5969,\allowbreak{}0.6511] and [0.5900,\allowbreak{}0.6476], respectively. These intervals are not error bars for the pooled pixel-level point estimate.\par

\begin{center}\begin{minipage}{\linewidth}\centering
\includegraphics[width=\linewidth,height=0.62\textheight,keepaspectratio]{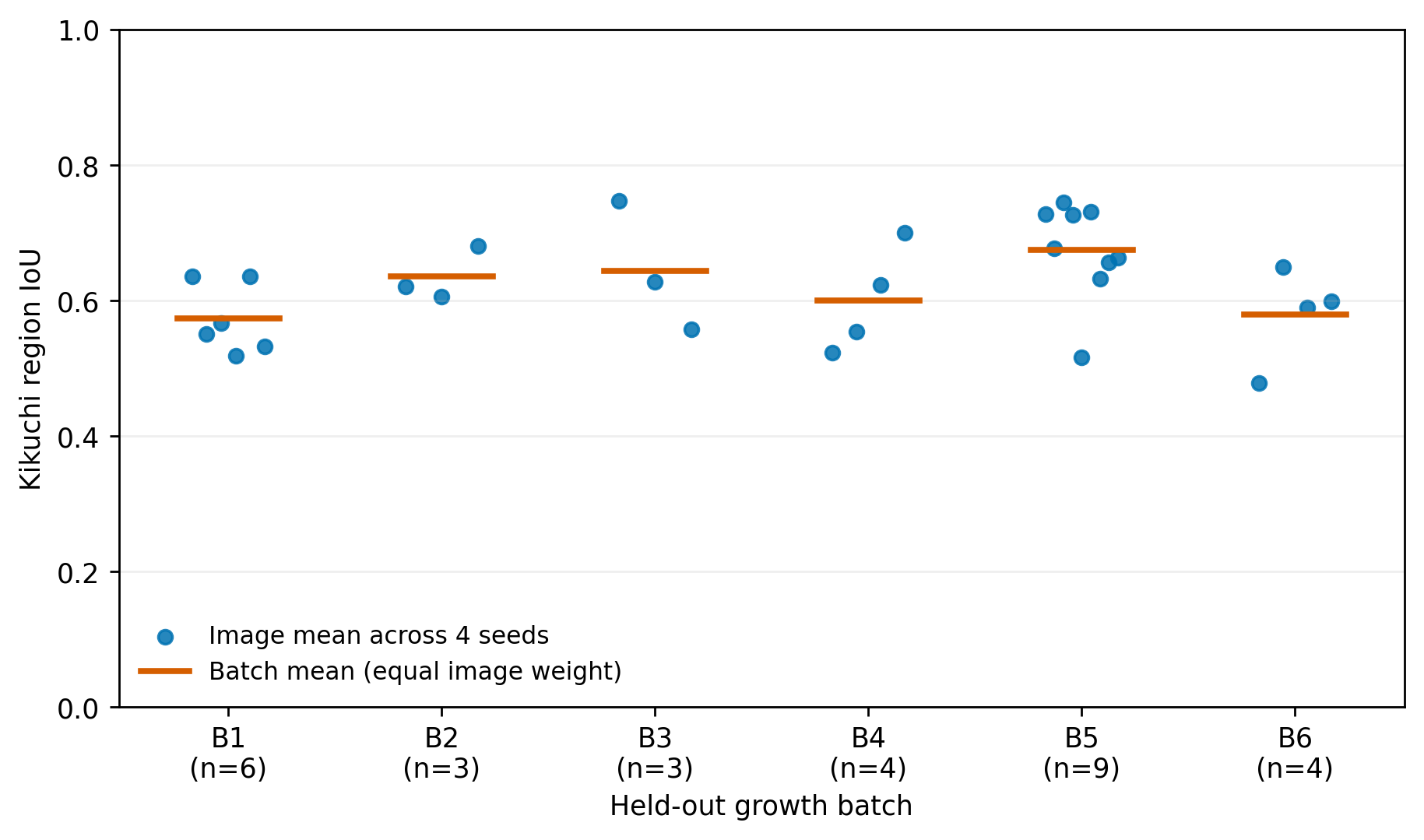}
\par\vspace{3mm}{\small\raggedright Figure S2. Internal per-image and per-batch results. Each point is the four-run mean IoU for one image; short lines show batch means over images. B1–B6 are ordered by original batch identifiers, not performance. All six batches are excluded from training and validation, with mean IoUs 0.5738–0.6756. Point dispersion reflects image differences rather than seed-error bars. B5 contains nine images; equal-batch weighting still gives an overall result close to equal-image weighting.\par}
\end{minipage}\end{center}

\subsubsection*{S4.2 Two Simple Processing Baselines}

\addcontentsline{toc}{subsubsection}{S4.2 Two Simple Processing Baselines}

Parameters were selected to maximize pooled IoU on the same 23 validation images and then fixed for the 29 test images. Grayscale percentile candidates were 50, 60, 70, 80, 85, 90, and 95, selecting the 70th percentile. Gaussian-residual candidates used \(\sigma=2,4,8,16\) pixels, positive or absolute residuals, and the same seven percentiles, selecting \(\sigma=16\), positive residuals, and the 85th percentile. Gaussian boundaries use reflection; there is no connected-region deletion or test-stage postprocessing. The searches examine 7 and 56 configurations, with ties resolved by the first candidate in the stated order.\par

Grayscale and residual baselines gave validation IoUs 0.2433/0.1569 and test IoUs 0.2308/0.1437. Test precision/recall were 0.2578/0.6879 and 0.2176/0.2972, respectively. They neither learn target semantics nor use diffraction geometry and are not reproductions of Ref. [30] or of all conventional RHEED analysis methods.\par

\subsubsection*{S4.3 Additional Negative Patterns}

\addcontentsline{toc}{subsubsection}{S4.3 Additional Negative Patterns}

We excluded batches corresponding to the primary internal dataset and two historical training inventories from the original laboratory archive, then sampled four images from each of the remaining ten batches, giving 40 candidates. Four observers assessed them independently using interfaces that hid models, masks, IoU, original filenames, batch information, and other observers' judgments. Ten images from five batches met the criterion: at least three negative judgments, including at least two non-developers; at most one judgment of unsuitability for this test; and no ambiguous or visible-feature judgment. Eligibility depended solely on human opinions. Eight images had three negative judgments and one unsuitability judgment; two had four negative judgments. “Unsuitable” did not mean positive.\par

\par\noindent\begin{minipage}{\linewidth}\small
Table S5. Internal negative-image responses at the fixed 0.45 threshold.\par\vspace{4pt}
\setlength{\tabcolsep}{3pt}
\renewcommand{\arraystretch}{1.2}
\begin{tabular}{@{}>{\raggedright\arraybackslash}p{\dimexpr 0.305882\linewidth-2\tabcolsep\relax}>{\raggedright\arraybackslash}p{\dimexpr 0.211765\linewidth-2\tabcolsep\relax}>{\raggedright\arraybackslash}p{\dimexpr 0.282353\linewidth-2\tabcolsep\relax}>{\raggedright\arraybackslash}p{\dimexpr 0.200000\linewidth-2\tabcolsep\relax}@{}}
\toprule
\textbf{Model seed} & \textbf{Images with any prediction} & \textbf{Pooled false-positive area} & \textbf{Median per-image area} \\
\midrule
123 & 8/10 & 0.394\% & 0.343\% \\
2026 & 10/10 & 2.957\% & 3.378\% \\
42 & 10/10 & 1.479\% & 1.200\% \\
7 & 10/10 & 0.445\% & 0.500\% \\
Mean scores of four models & 10/10 & 0.624\% & 0.470\% \\
\bottomrule
\end{tabular}
\end{minipage}\par

\begin{center}\begin{minipage}{\linewidth}\centering
\includegraphics[width=\linewidth,height=0.62\textheight,keepaspectratio]{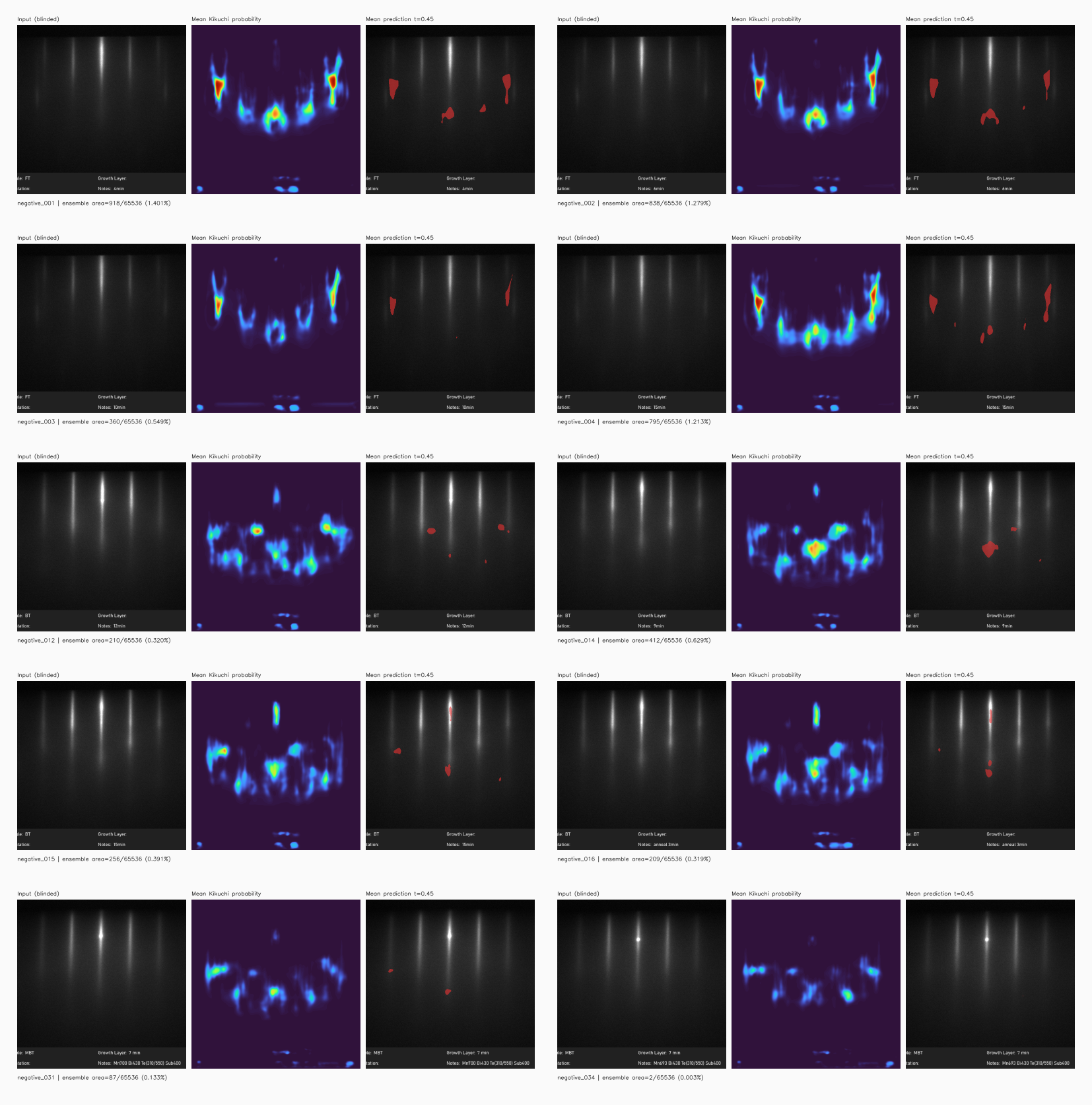}
\par\vspace{3mm}{\small\raggedright Figure S3. Each three-column group shows the input, four-model mean Kikuchi score, and region thresholded at 0.45; red marks extra predictions. Responses mainly occur at streak ends, central bright spots, and the lower screen edge rather than continuous oblique bands across the pattern. The ensemble is a supplementary illustration; each run is evaluated separately. Four of the ten images come from one batch. This small control does not estimate a general routine false-positive rate.\par}
\end{minipage}\end{center}

\subsection*{S5. Complete Public-Adaptation Curves, Resources, and Presentation Controls}

\addcontentsline{toc}{subsection}{S5. Complete Public-Adaptation Curves, Resources, and Presentation Controls}

\subsubsection*{S5.1 Primary Curves and Resources}

\addcontentsline{toc}{subsubsection}{S5.1 Primary Curves and Resources}

Twelve nested training sets use the same starting weights, 51 validation images, and 86 test images, with D4, AdamW, 24,000 updates, and validation every 400 updates. Seeds 20260931, 20260932, and 20260933 are integer identifiers, not dates. Smaller sets receive more presentations per image under the same budget. The zero-image evaluation and 15 completed runs at 20/\allowbreak{}40/\allowbreak{}100/\allowbreak{}160/\allowbreak{}244 images were reused after source-file hash verification. The 60/\allowbreak{}80/\allowbreak{}120/\allowbreak{}140/\allowbreak{}180/\allowbreak{}200/\allowbreak{}220 sizes add 21 new runs. Earlier 70-image results remain in the historical archive and are not included in the denser Figure 4 curve.\par

\par\noindent\begin{minipage}{\linewidth}\small
Table S6. Complete public-adaptation values. Three-reader grades map −1 to 1 before taking the median. Zero images refers to one starting model; other values are means±sample standard deviations. Grade 1 reports false-positive area rather than IoU.\par\vspace{4pt}
\setlength{\tabcolsep}{3pt}
\renewcommand{\arraystretch}{1.2}
\begin{tabular}{@{}>{\raggedright\arraybackslash}p{\dimexpr 0.164835\linewidth-2\tabcolsep\relax}>{\raggedright\arraybackslash}p{\dimexpr 0.142857\linewidth-2\tabcolsep\relax}>{\raggedright\arraybackslash}p{\dimexpr 0.142857\linewidth-2\tabcolsep\relax}>{\raggedright\arraybackslash}p{\dimexpr 0.142857\linewidth-2\tabcolsep\relax}>{\raggedright\arraybackslash}p{\dimexpr 0.142857\linewidth-2\tabcolsep\relax}>{\raggedright\arraybackslash}p{\dimexpr 0.263736\linewidth-2\tabcolsep\relax}@{}}
\toprule
\textbf{Training images} & \textbf{Overall IoU} & \textbf{Grades 4–5 IoU} & \textbf{Grade 3 IoU} & \textbf{Grade 2 IoU} & \textbf{Grade 1 false-positive area} \\
\midrule
0 & 0.1413 & 0.2040 & 0.1592 & 0.0662 & 0.790\% \\
20 & 0.3214±0.0119 & 0.4473±0.0062 & 0.4511±0.0165 & 0.0652±0.0152 & 1.024\% \\
40 & 0.3308±0.0131 & 0.4424±0.0276 & 0.4308±0.0026 & 0.1256±0.0125 & 0.929\% \\
60 & 0.3871±0.0079 & 0.4992±0.0116 & 0.4766±0.0092 & 0.1669±0.0184 & 0.584\% \\
80 & 0.4062±0.0163 & 0.5060±0.0224 & 0.4666±0.0132 & 0.1803±0.0128 & 0.246\% \\
100 & 0.4241±0.0040 & 0.5171±0.0073 & 0.4861±0.0124 & 0.1648±0.0094 & 0.107\% \\
120 & 0.4276±0.0148 & 0.5263±0.0136 & 0.4808±0.0314 & 0.1753±0.0174 & 0.101\% \\
140 & 0.4262±0.0188 & 0.5159±0.0144 & 0.4875±0.0323 & 0.1752±0.0179 & 0.077\% \\
160 & 0.4386±0.0199 & 0.5066±0.0155 & 0.5134±0.0361 & 0.1983±0.0264 & 0.071\% \\
180 & 0.4334±0.0142 & 0.5135±0.0224 & 0.5012±0.0190 & 0.1697±0.0195 & 0.058\% \\
200 & 0.4258±0.0126 & 0.5018±0.0221 & 0.4983±0.0127 & 0.1612±0.0053 & 0.047\% \\
220 & 0.4351±0.0263 & 0.5070±0.0374 & 0.5094±0.0266 & 0.1431±0.0284 & 0.015\% \\
244 & 0.4298±0.0290 & 0.5014±0.0336 & 0.4931±0.0426 & 0.1666±0.0224 & 0.023\% \\
\bottomrule
\end{tabular}
\end{minipage}\par

The 12 test sample groups were resampled in 5,000 paired bootstrap draws, sharing group draws across training sizes and paired seeds. Within each draw, pixel counts were combined separately for each model, followed by the mean of the three model IoUs. This captures source-group sampling uncertainty against fixed references, excluding re-annotation, repartitioning, and method selection. Overall IoU intervals for 20/\allowbreak{}40/\allowbreak{}60/\allowbreak{}80/\allowbreak{}100/\allowbreak{}160/\allowbreak{}220 minus 244 images were [−0.1441,\allowbreak{}−0.0655], [−0.1457,\allowbreak{}−0.0422], [−0.0878,\allowbreak{}−0.0059], [−0.0481,\allowbreak{}+0.0078], [−0.0243,\allowbreak{}+0.0172], [−0.0131,\allowbreak{}+0.0323], and [−0.0049,\allowbreak{}+0.0174], respectively. Intervals for 120/\allowbreak{}140/\allowbreak{}180/\allowbreak{}200 images are retained in the full statistical CSV in the local project archive. With 244 images, mean stratum F1 values were 0.6674 for grades 4–5, 0.6598 for grade 3, and 0.2851 for grade 2; overall F1 was 0.6008±0.0283. Corresponding \(FP\)/valid-area fractions were 2.02\%, 1.64\%, and 0.45\%, with 0.69\% overall. Fractions in positive strata are not negative-image false-positive rates. An average of 2.7 grade-1 images per run had any prediction.\par

\par\noindent\begin{minipage}{\linewidth}\small
Table S7. Image-resource accounting under the current reference conditions, without estimating wholly manual annotation time. Training composition uses individual grades 2–5 versus 1/−1, distinct from three-reader test stratification.\par\vspace{4pt}
\setlength{\tabcolsep}{3pt}
\renewcommand{\arraystretch}{1.2}
\begin{tabular}{@{}>{\raggedright\arraybackslash}p{\dimexpr 0.192308\linewidth-2\tabcolsep\relax}>{\raggedright\arraybackslash}p{\dimexpr 0.192308\linewidth-2\tabcolsep\relax}>{\raggedright\arraybackslash}p{\dimexpr 0.192308\linewidth-2\tabcolsep\relax}>{\raggedright\arraybackslash}p{\dimexpr 0.230769\linewidth-2\tabcolsep\relax}>{\raggedright\arraybackslash}p{\dimexpr 0.192308\linewidth-2\tabcolsep\relax}@{}}
\toprule
\textbf{Training images} & \textbf{Training groups covered} & \textbf{Training grades 2–5/1 or −1} & \textbf{Total including validation and test} & \textbf{Fraction of 381 images} \\
\midrule
20 & 20 & 14/6 & 157 & 41.2\% \\
40 & 32 & 30/10 & 177 & 46.5\% \\
60 & 37 & 45/15 & 197 & 51.7\% \\
80 & 38 & 59/21 & 217 & 57.0\% \\
100 & 38 & 71/29 & 237 & 62.2\% \\
120 & 38 & 87/33 & 257 & 67.5\% \\
140 & 38 & 99/41 & 277 & 72.7\% \\
160 & 38 & 113/47 & 297 & 78.0\% \\
180 & 38 & 124/56 & 317 & 83.2\% \\
200 & 38 & 136/64 & 337 & 88.5\% \\
220 & 38 & 144/76 & 357 & 93.7\% \\
244 & 38 & 153/91 & 381 & 100.0\% \\
\bottomrule
\end{tabular}
\end{minipage}\par

These resource counts include prediction-derived drafts from S1.2; the 157 images cannot all be described as independently manually confirmed.\par

\subsubsection*{S5.2 Public Training Process}

\addcontentsline{toc}{subsubsection}{S5.2 Public Training Process}

\begin{center}\begin{minipage}{\linewidth}\centering
\includegraphics[width=\linewidth,height=0.62\textheight,keepaspectratio]{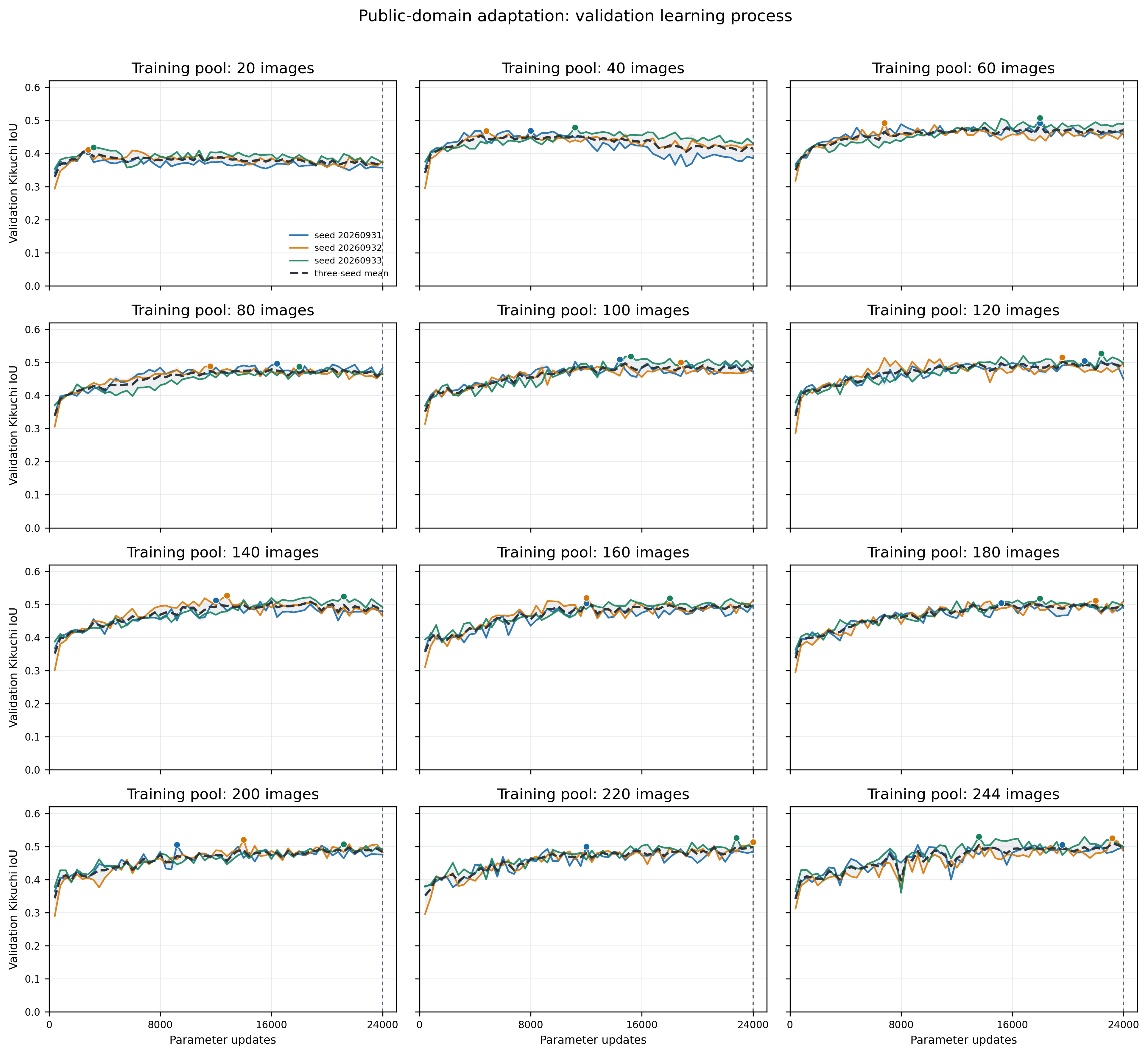}
\par\vspace{3mm}{\small\raggedright Figure S4. Validation Kikuchi IoU for three public-adaptation runs at each size. Validation uses the first author's references and grade-2–5 labels. Colored lines show individual runs, black dashed lines the mean, and gray bands the sample standard deviation. Circles mark the best-validation update; 24,000 updates is the common endpoint and need not be the saved checkpoint. The figure uses none of the 86 test images.\par}
\end{minipage}\end{center}

\begin{center}\begin{minipage}{\linewidth}\centering
\includegraphics[width=\linewidth,height=0.62\textheight,keepaspectratio]{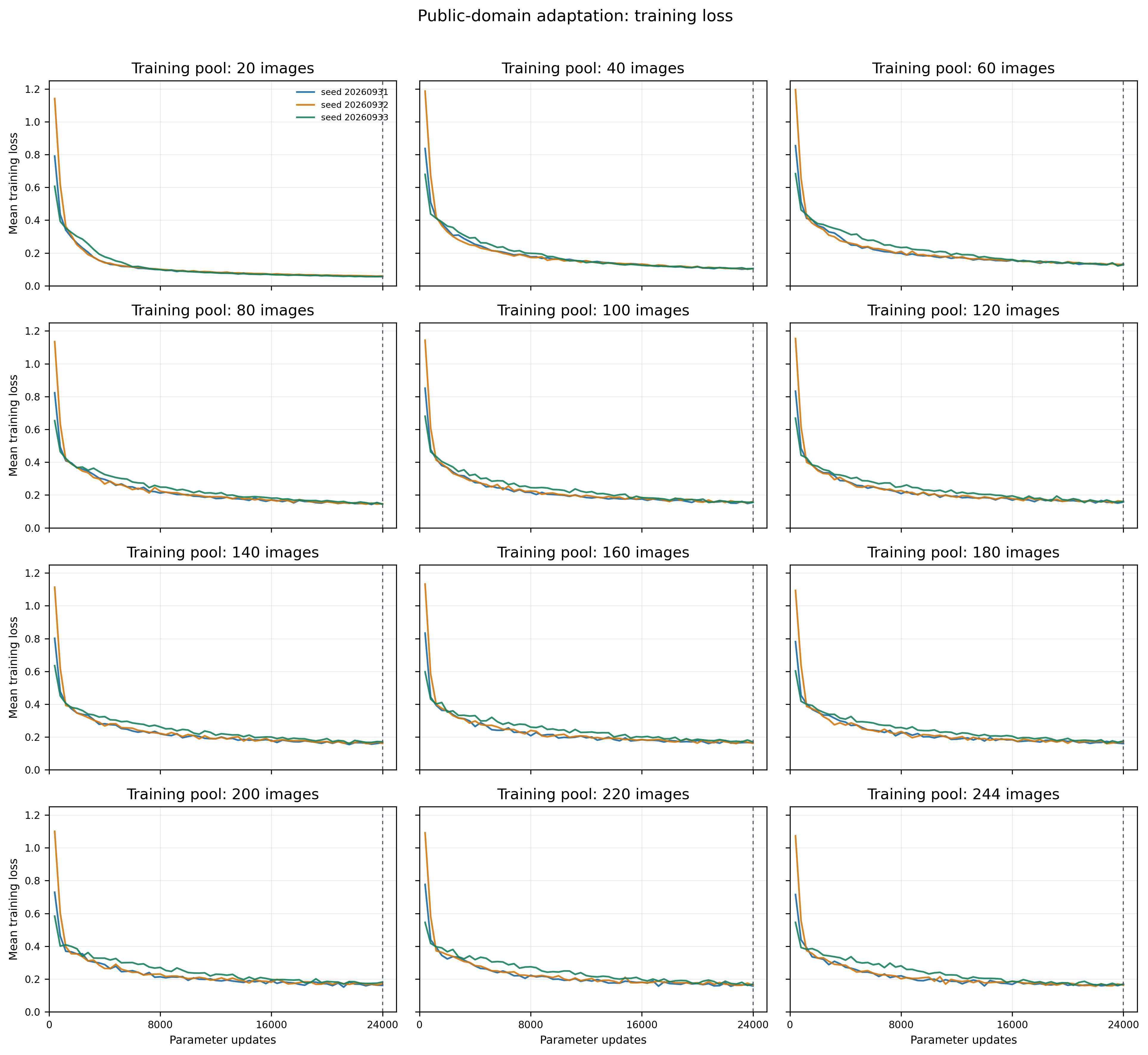}
\par\vspace{3mm}{\small\raggedright Figure S5. Public-adaptation training loss. Loss drives parameter optimization; checkpoint selection instead uses Kikuchi IoU on the fixed validation set. Presentations per image differ between training sizes.\par}
\end{minipage}\end{center}

\subsubsection*{S5.3 Matching Presentations per Image}

\addcontentsline{toc}{subsubsection}{S5.3 Matching Presentations per Image}

At 24,000 updates, 100 images receive an average of 240 presentations each, whereas 244 images receive 98.4. Additional runs with 100 images/9,840 updates and 244 images/58,560 updates form a 2×2 control. Initial parameters, training order, image roles, loss, D4, and seeds remain the same.\par

\par\noindent\begin{minipage}{\linewidth}\small
Table S8. Per-image presentation control, with means across three runs.\par\vspace{4pt}
\setlength{\tabcolsep}{3pt}
\renewcommand{\arraystretch}{1.2}
\begin{tabular}{@{}>{\raggedright\arraybackslash}p{\dimexpr 0.130435\linewidth-2\tabcolsep\relax}>{\raggedright\arraybackslash}p{\dimexpr 0.208696\linewidth-2\tabcolsep\relax}>{\raggedright\arraybackslash}p{\dimexpr 0.113043\linewidth-2\tabcolsep\relax}>{\raggedright\arraybackslash}p{\dimexpr 0.113043\linewidth-2\tabcolsep\relax}>{\raggedright\arraybackslash}p{\dimexpr 0.113043\linewidth-2\tabcolsep\relax}>{\raggedright\arraybackslash}p{\dimexpr 0.113043\linewidth-2\tabcolsep\relax}>{\raggedright\arraybackslash}p{\dimexpr 0.208696\linewidth-2\tabcolsep\relax}@{}}
\toprule
\textbf{Training images/updates} & \textbf{Mean presentations per image} & \textbf{Overall IoU} & \textbf{Grades 4–5 IoU} & \textbf{Grade 3 IoU} & \textbf{Grade 2 IoU} & \textbf{Grade 1 false-positive area} \\
\midrule
100/24,000 & 240.0 & 0.4241 & 0.5171 & 0.4861 & 0.1648 & 0.107\% \\
244/24,000 & 98.4 & 0.4298 & 0.5014 & 0.4931 & 0.1666 & 0.023\% \\
100/9,840 & 98.4 & 0.4144 & 0.5033 & 0.5014 & 0.1593 & 0.246\% \\
244/58,560 & 240.0 & 0.4396 & 0.5187 & 0.4999 & 0.1634 & 0.012\% \\
\bottomrule
\end{tabular}
\end{minipage}\par

At the lower presentation count, the 100-minus-244 difference is −0.0154, with 95\% interval [−0.0471,\allowbreak{}+0.0157]; at the higher count it is −0.0156, [−0.0399,\allowbreak{}+0.0105]. Extending the 244-image run adds 0.0098, [−0.0015,\allowbreak{}+0.0235]. All four conditions validate every 400 updates, giving approximately 25, 60, 60, and 147 candidate checkpoints. Longer runs also offer more checkpoint-selection opportunities, so the small increment cannot be attributed solely to presentations per image.\par

\begin{center}\begin{minipage}{\linewidth}\centering
\includegraphics[width=\linewidth,height=0.62\textheight,keepaspectratio]{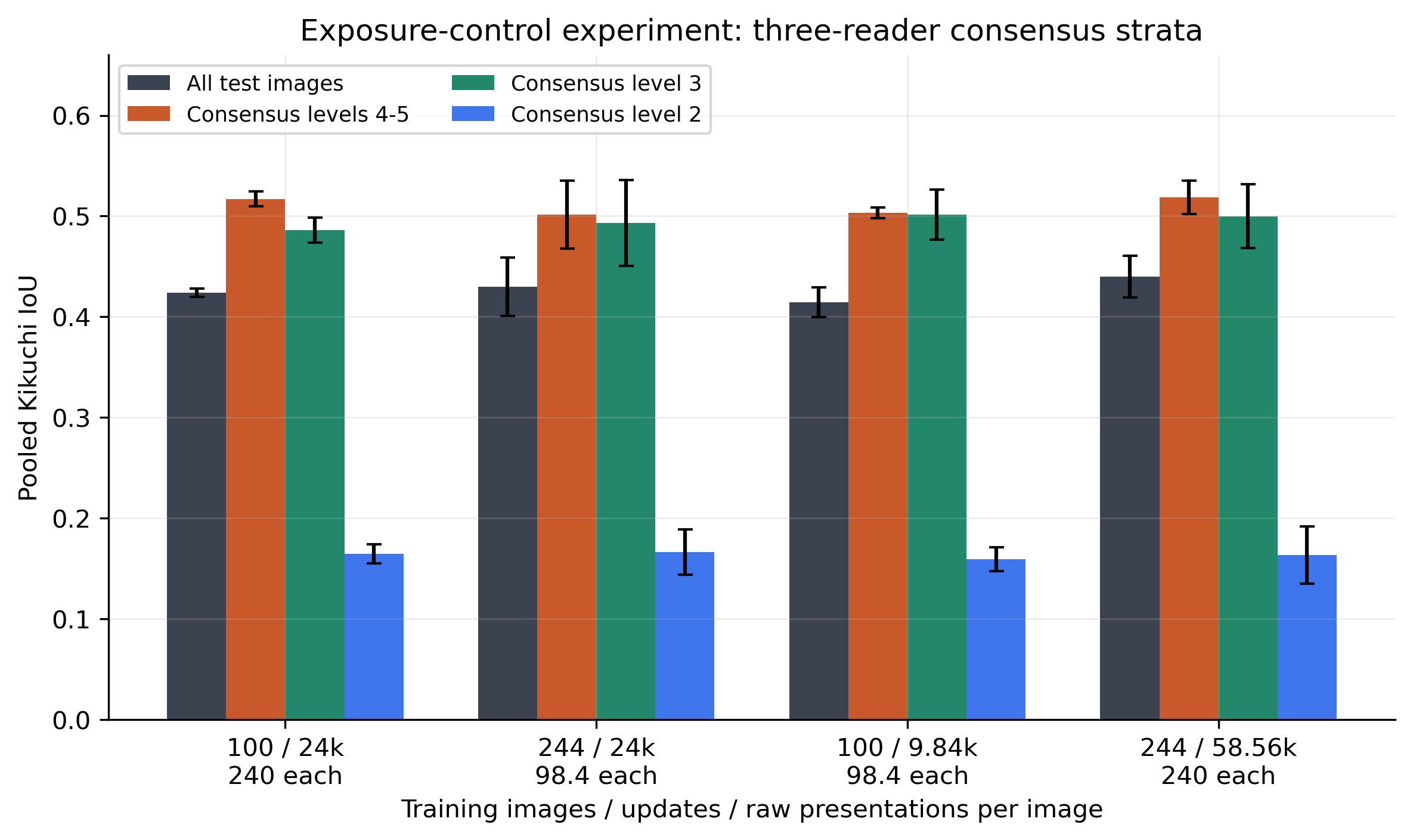}
\par\vspace{3mm}{\small\raggedright Figure S6. Overall and median-visibility-stratified results of the 2×2 control. Error bars show standard deviations across runs; horizontal labels specify training images, updates, and presentations per image.\par}
\end{minipage}\end{center}

\begin{center}\begin{minipage}{\linewidth}\centering
\includegraphics[width=\linewidth,height=0.62\textheight,keepaspectratio]{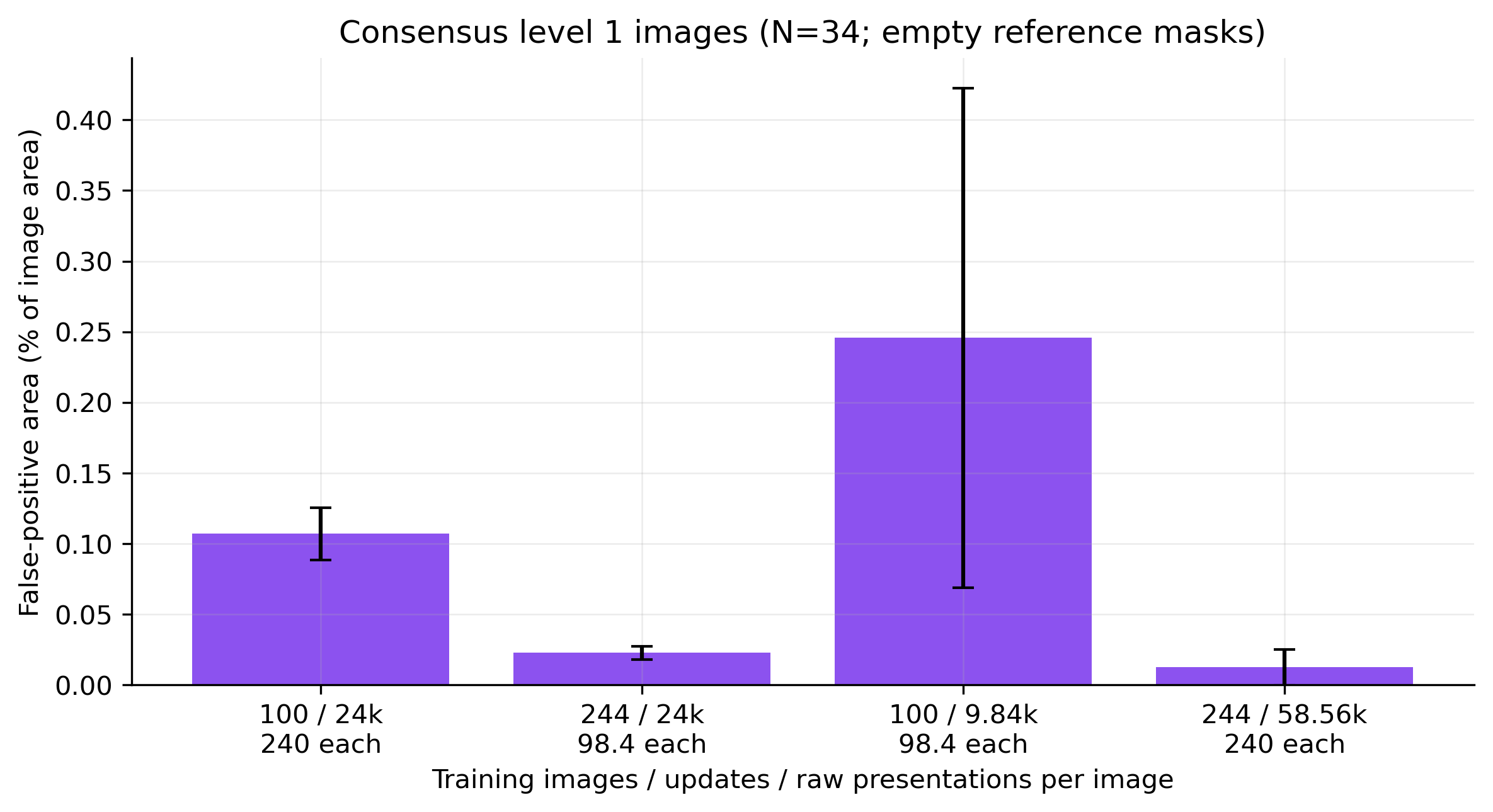}
\par\vspace{3mm}{\small\raggedright Figure S7. False-positive area for median grade 1 under the same conditions. Numerical trends are reported without an additional significance claim.\par}
\end{minipage}\end{center}

\subsection*{S6. Visibility Assessment, Observers, and Agreement}

\addcontentsline{toc}{subsection}{S6. Visibility Assessment, Observers, and Agreement}

\par\noindent\begin{minipage}{\linewidth}\small
Table S9. Observers, experience, and tasks. Abbreviations in parentheses identify archived records and reused figures rather than responsibilities in the main text.\par\vspace{4pt}
\setlength{\tabcolsep}{3pt}
\renewcommand{\arraystretch}{1.2}
\begin{tabular}{@{}>{\raggedright\arraybackslash}p{\dimexpr 0.121019\linewidth-2\tabcolsep\relax}>{\raggedright\arraybackslash}p{\dimexpr 0.292994\linewidth-2\tabcolsep\relax}>{\raggedright\arraybackslash}p{\dimexpr 0.292994\linewidth-2\tabcolsep\relax}>{\raggedright\arraybackslash}p{\dimexpr 0.292994\linewidth-2\tabcolsep\relax}@{}}
\toprule
\textbf{Observer} & \textbf{Author-confirmed MBE/RHEED experience} & \textbf{Task} & \textbf{Recorded dates and times (+08:00)} \\
\midrule
Lumou Weng (WLM) & Approximately two years of study and practice, including multiple independent MBE growth runs; model developer & Grading 382 images, region references, training-grade supervision & Grading: 09-12 13:11 to 09-13 20:17 \\
Hanshan Huang (HHS) & Approximately two years of study and practice, including multiple independent MBE growth runs & Grading 382 images; independent region re-annotation of 15 images & Grading: 09-13 17:58–18:46; re-annotation: 09-23 10:36–10:46 \\
Zhenhan Zhang (ZZH) & Approximately one year of study, with limited independent MBE growth experience & Grading 382 images & 09-13 20:19–21:59 \\
Gan Wang (WG) & Supervisor with extensive MBE/RHEED experience & Independent region re-annotation of 15 images & 09-23 10:50–10:56 \\
\bottomrule
\end{tabular}
\end{minipage}\par

All dates are in 2026. Prediction-blinding conditions for public grading and the three-reader aggregation rule are given in Section 3.3. Observers could increase display brightness, but per-image brightness values were not retained for historical public grading. Display adjustments do not enter model training or evaluation.\par

The interface for independent re-annotation of 15 images loaded no historical masks, models, streak labels, or other observers' annotations. Huang used brightness 0 for eight images and positive values for seven (maximum +38); Wang used 0 throughout. These records apply only to the 15-image re-annotation, not to grading of the 382 images.\par

\par\noindent\begin{minipage}{\linewidth}\small
Table S10. Ordinal agreement between observers on 382 public images, mapping −1 to 1.\par\vspace{4pt}
\setlength{\tabcolsep}{3pt}
\renewcommand{\arraystretch}{1.2}
\begin{tabular}{@{}>{\raggedright\arraybackslash}p{\dimexpr 0.195652\linewidth-2\tabcolsep\relax}>{\raggedright\arraybackslash}p{\dimexpr 0.184783\linewidth-2\tabcolsep\relax}>{\raggedright\arraybackslash}p{\dimexpr 0.195652\linewidth-2\tabcolsep\relax}>{\raggedright\arraybackslash}p{\dimexpr 0.163043\linewidth-2\tabcolsep\relax}>{\raggedright\arraybackslash}p{\dimexpr 0.260870\linewidth-2\tabcolsep\relax}@{}}
\toprule
\textbf{Observer pair} & \textbf{Exact agreement} & \textbf{Difference \(\leq\)1 grade} & \textbf{Linearly weighted kappa} & \textbf{Quadratically weighted kappa} \\
\midrule
Weng–Huang & 54.2\% & 93.5\% & 0.5965 & 0.7621 \\
Weng–Zhang & 40.1\% & 87.7\% & 0.4977 & 0.6892 \\
Huang–Zhang & 38.0\% & 85.9\% & 0.4175 & 0.5949 \\
\bottomrule
\end{tabular}
\end{minipage}\par

Counts for grades 1–5 were (152,88,63,49,30), (142,97,90,52,1), and (53,107,122,62,38) for Weng, Huang, and Zhang, respectively. Huang assigned grade 5 less often, while Zhang tended to assign higher grades, reflecting differences in category boundaries. Linear and quadratic kappa weight disagreements by grade difference or its square as sensitivity descriptions of ordinal agreement; the grades are not equally spaced physical quantities. Audits comparing mean and median rules were descriptive only; primary stratification consistently uses the median. For example, ratings 2/\allowbreak{}3/\allowbreak{}5 give grade 3. Labels −1 and 1 both indicate no target, while −1 retains its hard-negative identity during training.\par

\subsection*{S7. Independent Pixel Re-annotation and Linewidth Sensitivity}

\addcontentsline{toc}{subsection}{S7. Independent Pixel Re-annotation and Linewidth Sensitivity}

\subsubsection*{S7.1 Complete Statistics for 15 Re-annotated Images}

\addcontentsline{toc}{subsubsection}{S7.1 Complete Statistics for 15 Re-annotated Images}

All masks were placed on the 256×256 grid and model predictions thresholded at 0.45. Table S11 first pools pixel counts over the 15 images; these results do not replace the 29-image primary results. In a comparison “A versus B,” precision and recall treat A as prediction and B as reference; reversing the order exchanges them.\par

\par\noindent\begin{minipage}{\linewidth}\small
Table S11. Agreement between independent observers, original annotations, and model outputs.\par\vspace{4pt}
\setlength{\tabcolsep}{3pt}
\renewcommand{\arraystretch}{1.2}
\begin{tabular}{@{}>{\raggedright\arraybackslash}p{\dimexpr 0.300000\linewidth-2\tabcolsep\relax}>{\raggedright\arraybackslash}p{\dimexpr 0.162500\linewidth-2\tabcolsep\relax}>{\raggedright\arraybackslash}p{\dimexpr 0.162500\linewidth-2\tabcolsep\relax}>{\raggedright\arraybackslash}p{\dimexpr 0.212500\linewidth-2\tabcolsep\relax}>{\raggedright\arraybackslash}p{\dimexpr 0.162500\linewidth-2\tabcolsep\relax}@{}}
\toprule
\textbf{Comparison} & \textbf{IoU} & \textbf{Dice/F1} & \textbf{Precision} & \textbf{Recall} \\
\midrule
Huang versus Wang & 0.5516 & 0.7110 & 0.7023 & 0.7199 \\
Weng versus Huang & 0.4026 & 0.5741 & 0.8536 & 0.4325 \\
Weng versus Wang & 0.4102 & 0.5817 & 0.8509 & 0.4419 \\
Four models versus Huang & 0.3911–0.4093 & 0.5623–0.5808 & 0.8201–0.8576 & 0.4204–0.4496 \\
Four models versus Wang & 0.3968–0.4154 & 0.5651–0.5870 & 0.8157–0.8453 & 0.4210–0.4584 \\
\bottomrule
\end{tabular}
\end{minipage}\par

\begin{center}\begin{minipage}{\linewidth}\centering
\includegraphics[width=\linewidth,height=0.62\textheight,keepaspectratio]{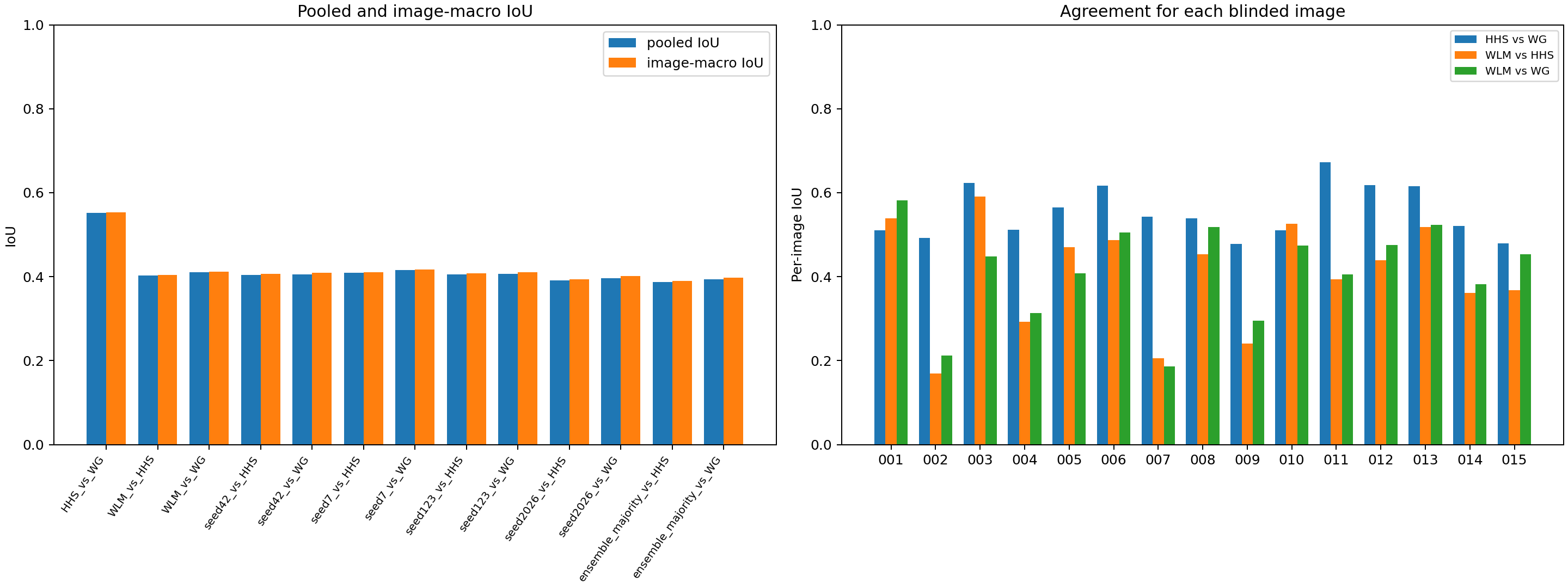}
\par\vspace{3mm}{\small\raggedright Figure S8. Left: pooled and equally weighted per-image IoU. Right: per-image distributions. WLM/HHS/WG correspond to Weng/Huang/Wang in Table S9. These comparisons explain boundary conventions and do not define a theoretical human-performance ceiling; full model ranges appear in Table S11.\par}
\end{minipage}\end{center}

Mean mask areas for Huang/Wang/Weng were 14,038.9/\allowbreak{}13,695.2/\allowbreak{}7,112.9 pixels; the four models gave 7,094.0/\allowbreak{}7,696.8/\allowbreak{}7,292.1/\allowbreak{}6,952.7 pixels for seeds 42/\allowbreak{}7/\allowbreak{}123/\allowbreak{}2026. Model outputs and original labels are narrower, with high precision and lower recall. Using the independent observers' intersection as a common core gives original-label IoU 0.4822 and model IoUs 0.4794–0.4895; using their union as a broader region gives 0.3591 and 0.3365–0.3662, respectively. Intersection and union are sensitivity references, not newly assigned ground truth. Human IoU across difficulty groups in these 15 images was not strictly monotonic, further distinguishing image visibility from boundary consistency.\par

\subsubsection*{S7.2 Reference Erosion with Fixed Outputs}

\addcontentsline{toc}{subsubsection}{S7.2 Reference Erosion with Fixed Outputs}

The four models' binary predictions at threshold 0.45 were fixed while reference masks were eroded inward by Euclidean distances of 0–10 pixels on the 256 grid. Each condition uses the same radius for all images. Erosion changes only the reference extent for analysis: originals, training, and thresholds remain fixed, and the highest-IoU eroded reference is not selected as a new primary reference.\par

\begin{center}\begin{minipage}{\linewidth}\centering
\includegraphics[width=\linewidth,height=0.62\textheight,keepaspectratio]{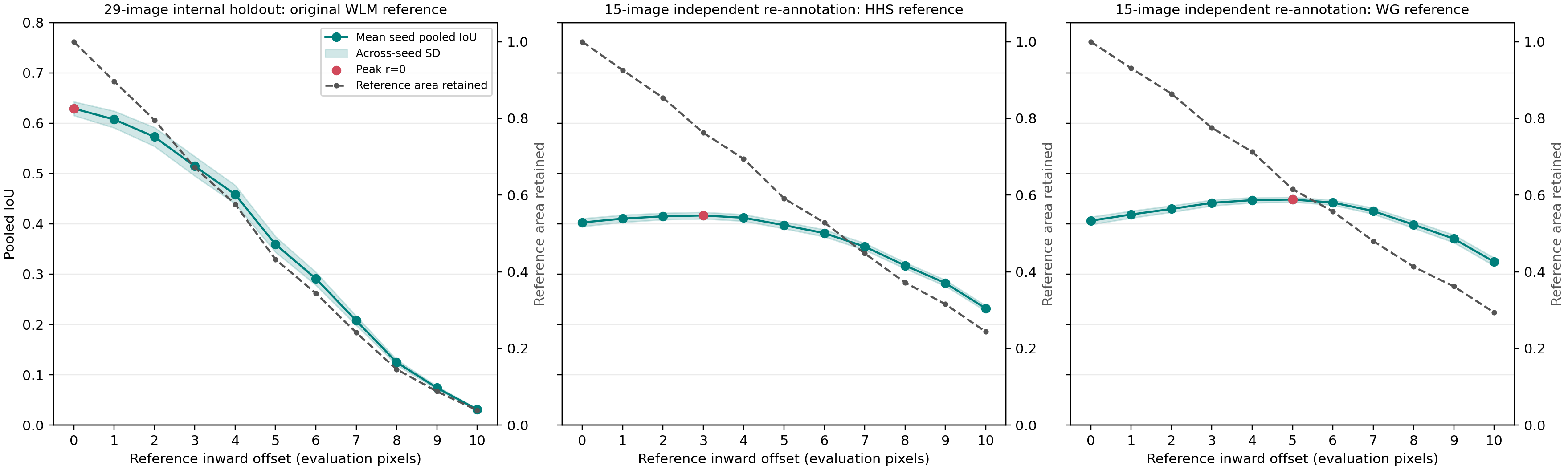}
\par\vspace{3mm}{\small\raggedright Figure S9. Linewidth sensitivity with fixed outputs. Solid lines show four-run mean pooled IoU, shading the standard deviation, gray dashed lines retained reference area, and red points post hoc maxima. Original references for 29 images are highest without erosion, at 0.6290±0.0137, and decrease monotonically thereafter. On 15 images, eroding Huang's references by 3 pixels retains 76.2\% of their area and raises IoU from 0.4027±0.0080 to 0.4166±0.0066. Eroding Wang's by 5 pixels retains 61.5\% and raises IoU from 0.4062±0.0076 to 0.4483±0.0048. Width explains only part of the differences; post hoc peaks do not replace scores against unmodified references.\par}
\end{minipage}\end{center}

\subsection*{S8. Six Design Controls under the Same Protocol}

\addcontentsline{toc}{subsection}{S8. Six Design Controls under the Same Protocol}

All controls share the full model's original labels, 93/\allowbreak{}23/\allowbreak{}29 batch partitions, 256 input, four paired seeds, maximum 80 epochs, checkpoint selection by validation F1 at 0.50, and testing at 0.45. Only the specified component changes. The D4 control repeats each original eight times to match views per epoch and nominal budget; actual stopping epochs and learning-rate trajectories remain in the logs. Each condition has 116 run–image records, with resampling by image or batch rather than treating four runs as independent test sets.\par

\par\noindent\begin{minipage}{\linewidth}\small
Table S12. Full-model-minus-control differences. Mean±training SD for pooled pixel-level differences and equal-batch differences with 95\% intervals are different estimands.\par\vspace{4pt}
\setlength{\tabcolsep}{3pt}
\renewcommand{\arraystretch}{1.2}
\begin{tabular}{@{}>{\raggedright\arraybackslash}p{\dimexpr 0.547619\linewidth-2\tabcolsep\relax}>{\raggedright\arraybackslash}p{\dimexpr 0.166667\linewidth-2\tabcolsep\relax}>{\raggedright\arraybackslash}p{\dimexpr 0.285714\linewidth-2\tabcolsep\relax}@{}}
\toprule
\textbf{Control modification} & \textbf{Pooled ΔIoU} & \textbf{Equal-batch ΔIoU [95\% interval]} \\
\midrule
Bypass the compressed-context branch & −0.0040±0.0144 & −0.0049 [−0.0227,\allowbreak{}+0.0125] \\
Zero encoder features in both skip concatenations & +0.0037±0.0291 & −0.0016 [−0.0125,\allowbreak{}+0.0114] \\
Replace independent outputs with four-state softmax & −0.0057±0.0086 & −0.0062 [−0.0138,\allowbreak{}+0.0007] \\
Remove auxiliary union BCE loss & +0.0042±0.0033 & +0.0001 [−0.0141,\allowbreak{}+0.0133] \\
Remove 0.2-weighted Kikuchi soft-Dice & +0.0056±0.0138 & +0.0074 [+0.0005,\allowbreak{}+0.0189] \\
Replace eight D4 views with eight original-image repetitions & +0.0091±0.0168 & +0.0081 [−0.0096,\allowbreak{}+0.0293] \\
\bottomrule
\end{tabular}
\end{minipage}\par

The four states are background, streak only, Kikuchi only, and overlap, so this control retains the ability to represent overlap with a different parameterization. Bypassing context retains the module but skips its forward calculation; other paths are unchanged. Removing union loss retains its head parameters. Context differences by seed were +0.0146/−0.0126/−0.0178/−0.0004; full-model mean IoU was 0.6290 versus 0.6330 for the control, with no consistent direction.\par

The soft-Dice equal-batch interval is positive, whereas the interval obtained by resampling batches and recombining pixels is [−0.0021,\allowbreak{}+0.0175]. D4 has the largest mean difference, but its interval includes zero. The controls thus offer limited directional evidence without establishing any added module as decisive or demonstrating model equivalence. All controls use 100,000 resamples.\par

\begin{center}\begin{minipage}{\linewidth}\centering
\includegraphics[width=\linewidth,height=0.62\textheight,keepaspectratio]{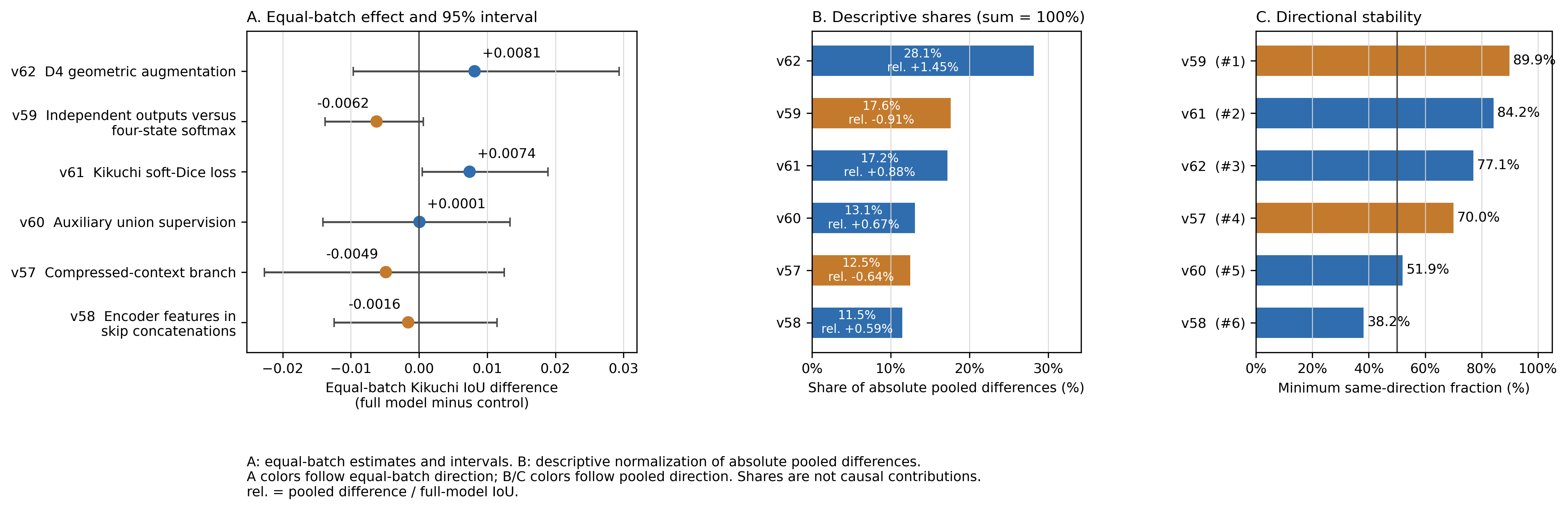}
\par\vspace{3mm}{\small\raggedright Figure S10. (A) Equal-batch point estimates and corresponding intervals. (B) Descriptive shares obtained by normalizing absolute pooled differences across the six controls. (C) The smallest same-direction fraction among three resampling schemes. Shares for D4, four-state output, soft-Dice, union, context, and skip connections are 28.1\%, 17.6\%, 17.2\%, 13.1\%, 12.5\%, and 11.5\%, respectively; these are not additive causal contributions. Colors in A follow equal-batch direction, while B/C use pooled direction. Same-direction fractions are not probabilities of mechanism effectiveness. All positive and negative results are retained.\par}
\end{minipage}\end{center}

\subsection*{S9. Region References and Predictions for Fixed Illustrations}

\addcontentsline{toc}{subsection}{S9. Region References and Predictions for Fixed Illustrations}

\begin{center}\begin{minipage}{\linewidth}\centering
\includegraphics[width=\linewidth,height=0.62\textheight,keepaspectratio]{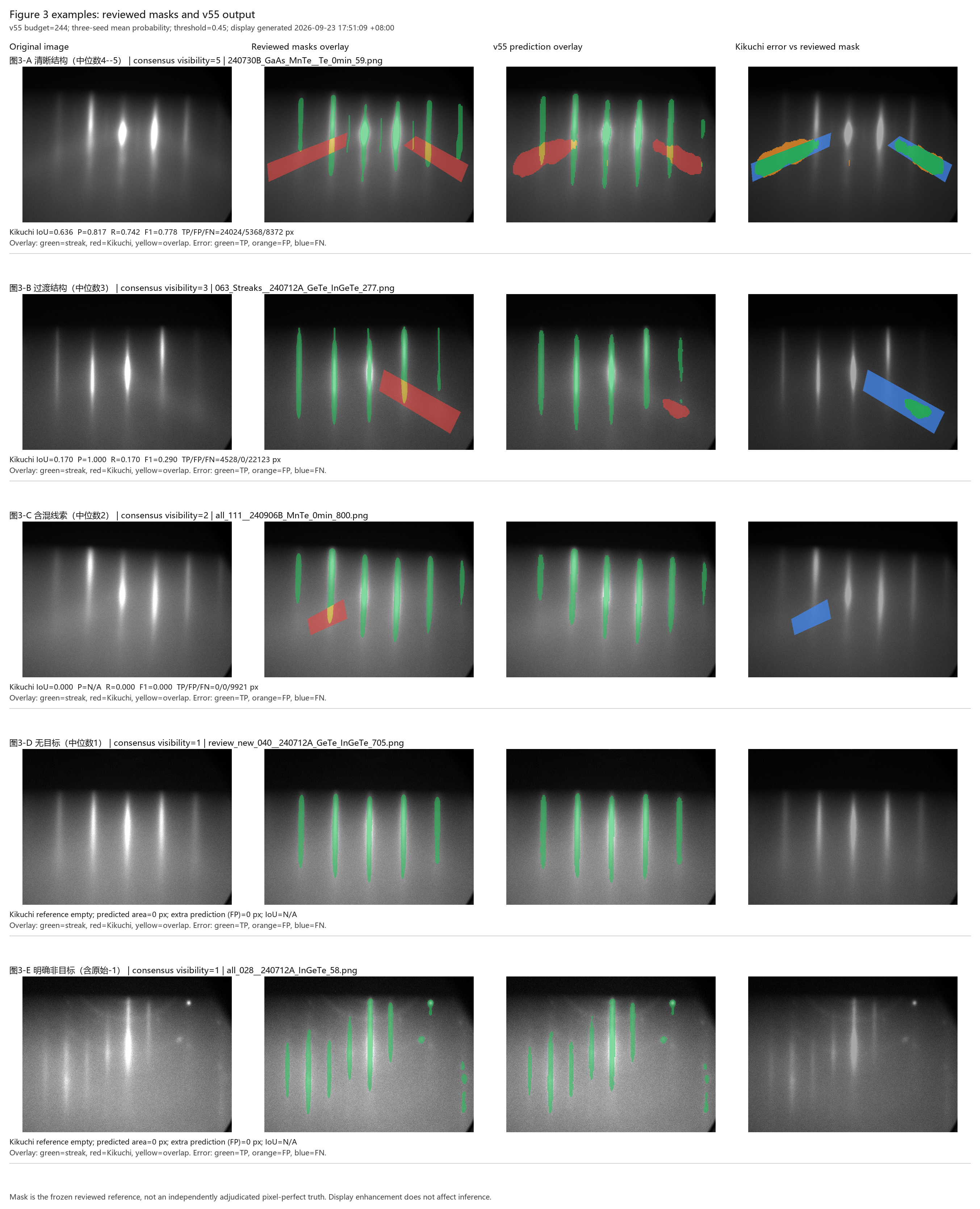}
\par\vspace{3mm}{\small\raggedright Figure S11. Complete comparisons for the same five images as Figure 3, selected without IoU. Rows show the displayed original, fixed two-class reference, mean-score prediction from three models trained on all 244 images, and Kikuchi errors. Scores are averaged on the 256 grid, thresholded at 0.45, unpadded, and restored to original size. The 1–99.5 percentile display stretch does not enter inference. Overlays use green for streaks, red for Kikuchi regions, and yellow for overlap; error maps use green/orange/blue for correct/extra/missed pixels. D/E have empty references and predictions, zero extra-response pixels, and NA IoU. Ensemble illustrations differ from the separate-run statistics in Table S6; prediction-draft provenance is described in S1.2.\par}
\end{minipage}\end{center}

\begin{center}\begin{minipage}{\linewidth}\centering
\includegraphics[width=\linewidth,height=0.62\textheight,keepaspectratio]{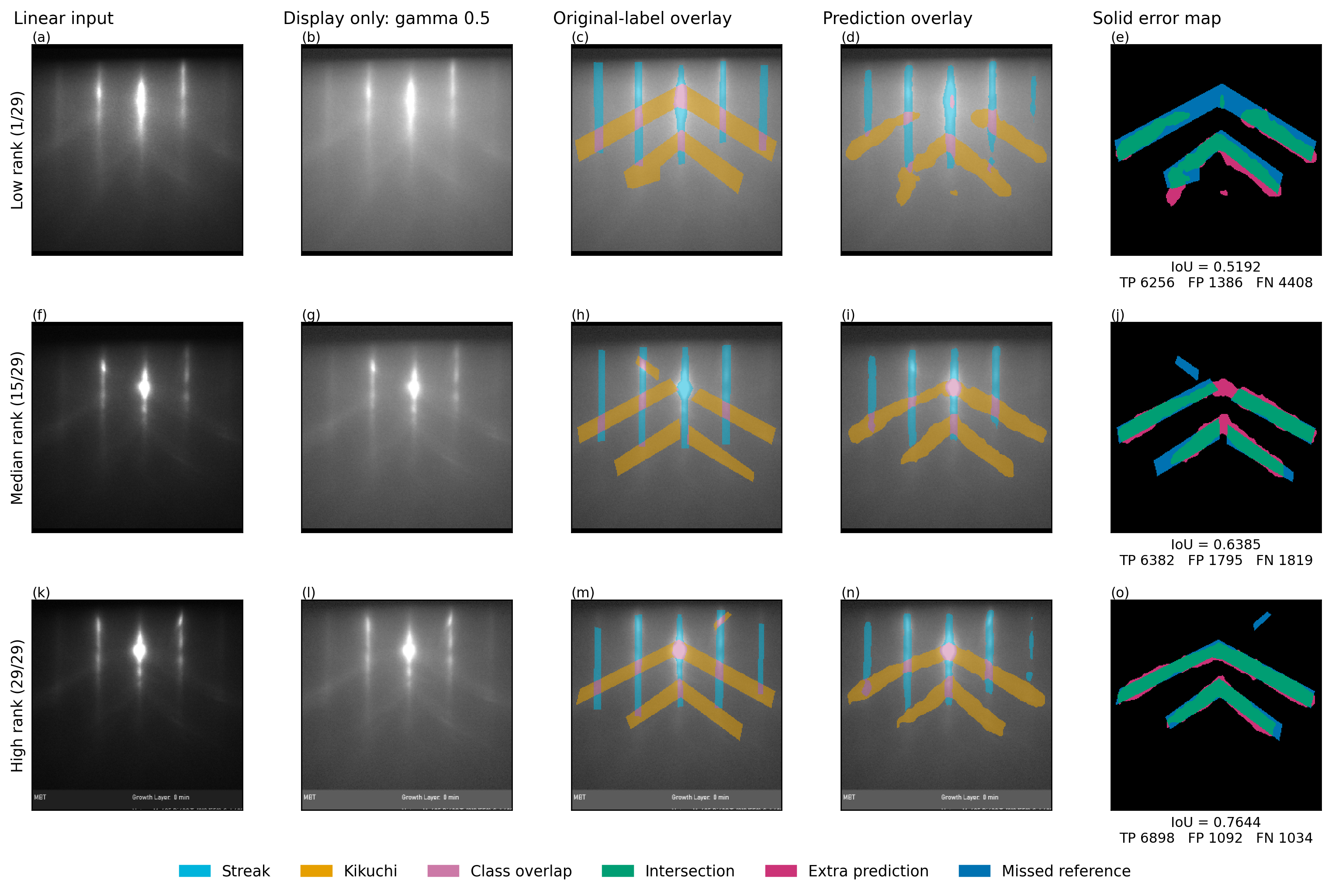}
\par\vspace{3mm}{\small\raggedright Figure S12. The 29 images are ranked by four-run mean IoU, and ranks 1, 15, and 29 are shown with seed-42 predictions throughout. Columns are linear display, square-root display, original two-class reference, prediction, and Kikuchi error. Model IoUs are 0.5192, 0.6385, and 0.7644, respectively. The low-ranked example has 6,256 correctly covered and 4,408 missed pixels. Overlay and error colors match Figure 1, illustrating weak-line omissions, discontinuity, and boundary displacement. References do not include revisions based on model outputs.\par}
\end{minipage}\end{center}

\subsection*{S10. Organization of Reproducibility Materials}

\addcontentsline{toc}{subsection}{S10. Organization of Reproducibility Materials}

Materials in the local project archive are organized by primary internal training with original labels, public adaptation, presentation controls, independent re-annotation, negative controls, principal design controls, and figures. Each retains input and partition manifests, model identity, training settings, validation checkpoint rules, thresholds, evaluation grids, per-image \(TP/FP/FN/TN\), summaries, and figure-generation scripts. Public-draft provenance, pending-review status, and individual-grade supervision are also retained. Descriptive names in the paper identify tasks; original program filenames and version mappings are recorded in the local reproducibility inventory. The materials remain locally archived as stated in the Data, Code, and AI-Assistance Statement; S12 documents the computational procedures included in this preprint.\par

\subsection*{S11. Computer-Vision Terms for Physics Readers}

\addcontentsline{toc}{subsection}{S11. Computer-Vision Terms for Physics Readers}

The main text introduces image representation, network paths, and evaluation. This section explains terms that can be confused with physical concepts. Exact structures and numerical rules appear in Figure 2, S1–S2, and Algorithms S1–S6.\par

\subsubsection*{S11.1 Pixels, Grayscale, Masks, and Feature Channels}

\addcontentsline{toc}{subsubsection}{S11.1 Pixels, Grayscale, Masks, and Feature Channels}

A pixel is one sampling position on a grid. A 256×256 grayscale image has 65,536 brightness values; a color image usually stores red, green, and blue components at each position. A mask records target positions, not diffraction intensity; its definition and overlapping representation appear in Section 2.1.\par

A feature map is a response map computed by the network, and channels are different responses on the same grid. One grayscale input can produce 48 feature channels describing morphological cues such as edges, orientation, or broader brightness changes. They are learned responses, not 48 acquisitions or detectors.\par

\subsubsection*{S11.2 Convolution, Convolution Blocks, and Nonlinear Responses}

\addcontentsline{toc}{subsubsection}{S11.2 Convolution, Convolution Blocks, and Nonlinear Responses}

Convolution uses the same learned coefficients at every position to combine neighboring values; multichannel convolution also combines responses from different channels. ReLU, introduced in Section 2.2, retains positive values and removes negative values, making successive multilayer computation more expressive than one fixed linear transformation.\par

A 1×1 convolution combines channels at the same position without enlarging the spatial window. Here it compresses and restores channels and generates output scores. The two 3×3 convolutions in DC and their dimensions appear in Figure 2 and Algorithm S2.\par

\subsubsection*{S11.3 Pooling, Encoder, Reconstruction, and Skip Connections}

\addcontentsline{toc}{subsubsection}{S11.3 Pooling, Encoder, Reconstruction, and Skip Connections}

Pooling summarizes a window by one number: max pooling takes its maximum and average pooling its mean. The encoder progressively combines information on smaller grids. A receptive field is the area of the original image that can influence a response; “deep” refers to computational layers, not physical depth in a specimen. The bottleneck is the smallest spatial grid in the main path, here 64×64.\par

Reconstruction increases grid density through interpolation; skip connections also restore spatial detail from shallower layers. Bilinear interpolation estimates new values from neighbors, while nearest-neighbor interpolation takes the closest value directly. Their uses are specified in S1.3. Concatenation retains two sets of channels side by side for subsequent convolution, rather than adding them point by point. Figure 2 shows the complete grid sequence.\par

\subsubsection*{S11.4 Vectors, Normalization, Context, Self-Attention, and Residuals}

\addcontentsline{toc}{subsubsection}{S11.4 Vectors, Normalization, Context, Self-Attention, and Residuals}

A vector is an ordered group of numbers. The compressed branch's 8×8 grid has 64 positions with 64 feature components each, forming 64-dimensional vectors. Position count and vector dimension happen to be equal but have different meanings. “Projection” means a learned linear combination, rather than geometrical projection of diffraction onto the screen.\par

LayerNorm adjusts components using the mean and variance within each position's vector, followed by learned scaling and shifting. It acts on internal responses, distinct from normalizing input grayscale by division by 255.\par

Context means image information over a wider area. Self-attention uses numerical representations \(Q\), \(K\), and \(V\) to compute positional associations and combine responses by weights (S2.1), without prescribing crystallographic planes. “Single head” means one set of association weights. A residual connection adds a branch result to the original response, with its magnitude controlled here by the learned \(\gamma\). Dropout2d temporarily zeros some entire channels at random during training and is disabled for inference. Parameters and controls appear in S2.1 and S8.\par

\subsubsection*{S11.5 Output Heads, Logits, Sigmoid, Thresholds, and Overlap}

\addcontentsline{toc}{subsubsection}{S11.5 Output Heads, Logits, Sigmoid, Thresholds, and Overlap}

An output head is the final computation generating a specified result. Sigmoid converts logits to 0–1 scores (Section 2.2), which are not calibrated probabilities of correctness. A threshold converts scores to binary regions: raising it usually reduces region extent, and lowering it enlarges regions, changing the balance of omissions and extra predictions. The two class scores are processed independently; thresholds appear in Algorithms S3–S5.\par

Softmax converts a set of scores to weights summing to 1. The four-state control in S8 chooses among background, streak only, Kikuchi only, and overlap. It can therefore represent overlap but uses different parameterization from two independent sigmoid outputs.\par

\subsubsection*{S11.6 Supervision, Loss, Gradients, Optimization, and Augmentation}

\addcontentsline{toc}{subsubsection}{S11.6 Supervision, Loss, Gradients, Optimization, and Augmentation}

Supervised learning adjusts parameters using annotated examples. Loss defines discrepancies to minimize, whereas IoU evaluates overlap after thresholding; their roles differ (Sections 2.3–2.4). Backpropagation computes parameter gradients layer by layer from the final loss, and the optimizer updates parameters accordingly. Learning rate and weight decay are explained in Section 3.2.\par

A batch is the group of images processed in one update; an epoch is one complete pass through the training data. Augmentation changes views seen by the model without adding independent originals and must transform masks identically. The eight D4 orientations and update-budget-matched control are described in Section 3.2 and S8.\par

\subsubsection*{S11.7 Validation Selection, Fixed Evaluation, Seeds, and Uncertainty}

\addcontentsline{toc}{subsubsection}{S11.7 Validation Selection, Fixed Evaluation, Seeds, and Uncertainty}

Training data adjust parameters, validation data select checkpoints, and test data evaluate selected models. Here, batch- or sample-group separation keeps similar images from the same source in one partition. Evaluation after fixing data, references, model, and threshold is termed fixed evaluation.\par

A random seed is an integer controlling processes such as initialization and training order. Standard deviation across seeds describes training variability; intervals obtained by resampling images or complete source groups describe uncertainty from finite test sources. Pooled, equally weighted image, and equally weighted batch statistics are defined in Section 2.4, with complete procedures in S12.6.\par

\subsection*{S12. Reproducibility Pseudocode: From Images to Training and Evaluation}

\addcontentsline{toc}{subsection}{S12. Reproducibility Pseudocode: From Images to Training and Evaluation}

Algorithms S1–S2 give shared computations, S3 internal training, S4 public adaptation, and S5–S6 inference and statistics. Reproduction should read the fixed manifests and references listed in S10 without repartitioning based on test scores. Input masks align with the originals, and class order is fixed as streak, Kikuchi. Before starting, check file completeness, binary values, and source-group separation. The workflow contains no brightness augmentation, two-input fusion, test-time orientation augmentation, or morphological boundary refinement.\par

\subsubsection*{S12.1 Image and Annotation Preprocessing}

\addcontentsline{toc}{subsubsection}{S12.1 Image and Annotation Preprocessing}

\texttt{Resize} denotes interpolated resizing, \texttt{Pad} zero padding, and \texttt{geometry} the dimension and padding record. Interpolation and internal-entry consistency are discussed in S1.3. Public training masks use a strict greater-than-0.5 test after resizing.\par

\par\needspace{5\baselineskip}\begingroup\small\ttfamily\raggedright
\setlength{\parindent}{0pt}\setlength{\parskip}{1pt}
\noindent\hangindent=2em\hangafter=1\hspace*{0.0em}Algorithm S1  PREPARE(record,\allowbreak{} experiment)\par
\noindent\hangindent=2em\hangafter=1\hspace*{0.0em}Input: original image I; independent masks Gs,\allowbreak{} Gk; experiment internal or public\par
\noindent\hangindent=2em\hangafter=1\hspace*{0.0em}Output: 256-grid X,\allowbreak{} training labels Ys/Yk,\allowbreak{} original-size Gs/Gk,\allowbreak{} geometry\par
\par\vspace{0.3\baselineskip}
\noindent\hangindent=2em\hangafter=1\hspace*{0.0em}if experiment == internal:\par
\noindent\hangindent=2em\hangafter=1\hspace*{2.0em}I8 = ReadGray8(I)\par
\noindent\hangindent=2em\hangafter=1\hspace*{2.0em}X = Float32(Resize(I8,\allowbreak{} 256,\allowbreak{} 256,\allowbreak{} bilinear)) / 255\par
\noindent\hangindent=2em\hangafter=1\hspace*{2.0em}Ys = Resize(Gs,\allowbreak{} 256,\allowbreak{} 256,\allowbreak{} nearest) > 0\par
\noindent\hangindent=2em\hangafter=1\hspace*{2.0em}Yk = Resize(Gk,\allowbreak{} 256,\allowbreak{} 256,\allowbreak{} nearest) > 0\par
\noindent\hangindent=2em\hangafter=1\hspace*{2.0em}geometry = original dimensions and direct-resize record\par
\noindent\hangindent=2em\hangafter=1\hspace*{0.0em}else:\par
\noindent\hangindent=2em\hangafter=1\hspace*{2.0em}X0 = ReadPreservingDtype(I)\par
\noindent\hangindent=2em\hangafter=1\hspace*{2.0em}X0 = grayscale float values / maximum specified by data type\par
\noindent\hangindent=2em\hangafter=1\hspace*{2.0em}X0 = Clip(X0,\allowbreak{} 0,\allowbreak{} 1)\par
\noindent\hangindent=2em\hangafter=1\hspace*{2.0em}r = min(256 / original\_width,\allowbreak{} 256 / original\_height)\par
\noindent\hangindent=2em\hangafter=1\hspace*{2.0em}W1 = max(1,\allowbreak{} round(original\_width*r))\par
\noindent\hangindent=2em\hangafter=1\hspace*{2.0em}H1 = max(1,\allowbreak{} round(original\_height*r))\par
\noindent\hangindent=2em\hangafter=1\hspace*{2.0em}left,\allowbreak{} top = floor((256-W1)/2),\allowbreak{} floor((256-H1)/2)\par
\noindent\hangindent=2em\hangafter=1\hspace*{2.0em}X = Pad(Resize(X0,\allowbreak{} W1,\allowbreak{} H1,\allowbreak{} area),\allowbreak{} left,\allowbreak{} top,\allowbreak{} 256,\allowbreak{} 256,\allowbreak{} value=0)\par
\noindent\hangindent=2em\hangafter=1\hspace*{2.0em}for G in [Gs,\allowbreak{} Gk]:\par
\noindent\hangindent=2em\hangafter=1\hspace*{4.0em}Y = Pad(Resize(Float32(G > 0),\allowbreak{} W1,\allowbreak{} H1,\allowbreak{} area),\allowbreak{} same geometry,\allowbreak{} value=0)\par
\noindent\hangindent=2em\hangafter=1\hspace*{4.0em}corresponding training label = Float32(Y > 0.5)\par
\noindent\hangindent=2em\hangafter=1\hspace*{2.0em}geometry = original dimensions,\allowbreak{} W1/H1,\allowbreak{} and four padding amounts\par
\noindent\hangindent=2em\hangafter=1\hspace*{0.0em}return X,\allowbreak{} Float32(Ys),\allowbreak{} Float32(Yk),\allowbreak{} Gs,\allowbreak{} Gk,\allowbreak{} geometry\par
\endgroup\par\medskip

The data-type maxima are 255 for 8-bit and 65535 for 16-bit input. Floating-point inputs are treated as already normalized and clipped to 0–1, without per-image maximum or percentile normalization. Padding contributes to training loss as background; evaluation extent is restored by Algorithm S5.\par

D4 acts synchronously on X, Ys, and Yk. The eight modes are the original, horizontal reflection, vertical reflection, 180° rotation, clockwise 90° rotation, counterclockwise 90° rotation, transposition, and transposition followed by 180° rotation. Traversal is specified in Algorithms S3 and S4.\par

\subsubsection*{S12.2 Two-Class Localization and Auxiliary Visibility Calculation}

\addcontentsline{toc}{subsubsection}{S12.2 Two-Class Localization and Auxiliary Visibility Calculation}

Algorithm S2 omits the batch dimension; shapes are height×width×channels, and \texttt{Concat} concatenates channels. \texttt{DC} and \texttt{Up} follow Figure 2 and S2.1.\par

\par\needspace{5\baselineskip}\begingroup\small\ttfamily\raggedright
\setlength{\parindent}{0pt}\setlength{\parskip}{1pt}
\noindent\hangindent=2em\hangafter=1\hspace*{0.0em}Algorithm S2  FORWARD(X,\allowbreak{} public\_auxiliary,\allowbreak{} training)\par
\noindent\hangindent=2em\hangafter=1\hspace*{0.0em}E1 = DC(1->48,\allowbreak{} X)                         \# 256*256*48\par
\noindent\hangindent=2em\hangafter=1\hspace*{0.0em}E2 = DC(48->96,\allowbreak{} MaxPool2(E1))             \# 128*128*96\par
\noindent\hangindent=2em\hangafter=1\hspace*{0.0em}B  = DC(96->192,\allowbreak{} MaxPool2(E2))            \# 64*64*192\par
\noindent\hangindent=2em\hangafter=1\hspace*{0.0em}T  = Conv1x1(192->64,\allowbreak{} AvgPool8(B))        \# 8*8*64\par
\noindent\hangindent=2em\hangafter=1\hspace*{0.0em}Z  = LayerNorm(ReshapePositions(T))       \# 64 positions,\allowbreak{} 64 components each\par
\noindent\hangindent=2em\hangafter=1\hspace*{0.0em}Q,\allowbreak{} K,\allowbreak{} V = three learned linear projections of Z,\allowbreak{} with biases\par
\noindent\hangindent=2em\hangafter=1\hspace*{0.0em}A  = RowSoftmax(Q * Transpose(K) / sqrt(64))\par
\noindent\hangindent=2em\hangafter=1\hspace*{0.0em}Y  = learned linear output projection of A * V,\allowbreak{} with bias\par
\noindent\hangindent=2em\hangafter=1\hspace*{0.0em}C  = Up8(Conv1x1(64->192,\allowbreak{} ReshapeGrid(Y)))\par
\noindent\hangindent=2em\hangafter=1\hspace*{0.0em}B  = B + gamma*C                         \# trainable gamma,\allowbreak{} initialized to 0\par
\noindent\hangindent=2em\hangafter=1\hspace*{0.0em}B  = Dropout2d(B,\allowbreak{} probability=0.05) if training else B\par
\noindent\hangindent=2em\hangafter=1\hspace*{0.0em}D2 = DC(288->96,\allowbreak{} Concat(Up2(B),\allowbreak{} E2))\par
\noindent\hangindent=2em\hangafter=1\hspace*{0.0em}D1 = DC(144->48,\allowbreak{} Concat(Up2(D2),\allowbreak{} E1))\par
\noindent\hangindent=2em\hangafter=1\hspace*{0.0em}zu = Conv1x1(48->1,\allowbreak{} D1)\par
\noindent\hangindent=2em\hangafter=1\hspace*{0.0em}zs,\allowbreak{} zk = SplitChannels(Conv1x1(48->2,\allowbreak{} D1))\par
\noindent\hangindent=2em\hangafter=1\hspace*{0.0em}if public\_auxiliary:\par
\noindent\hangindent=2em\hangafter=1\hspace*{2.0em}v = [SpatialMean(zu),\allowbreak{} SpatialMean(zs),\allowbreak{} SpatialMean(zk)]\par
\noindent\hangindent=2em\hangafter=1\hspace*{2.0em}q = Linear(24->1,\allowbreak{} ReLU(Linear(3->24,\allowbreak{} v)))\par
\noindent\hangindent=2em\hangafter=1\hspace*{2.0em}return zu,\allowbreak{} zs,\allowbreak{} zk,\allowbreak{} q\par
\noindent\hangindent=2em\hangafter=1\hspace*{0.0em}return zu,\allowbreak{} zs,\allowbreak{} zk\par
\endgroup\par\medskip

The auxiliary head takes spatial means of logits; its output is explained in Section 3.3. Attention has no positional encoding, and LayerNorm uses a stabilizing constant of \(10^{-5}\). Convolutions, linear layers, and the attention output projection use biases. The auxiliary union head is distinct from the logical union of the two class predictions (S12.5).\par

\subsubsection*{S12.3 Independent Training with Original Laboratory Annotations}

\addcontentsline{toc}{subsubsection}{S12.3 Independent Training with Original Laboratory Annotations}

\texttt{BCE(z,G)} receives logits directly and averages over all pixels; \texttt{DiceLoss} is \(L_{D,k}\) in the main text. GPU training uses automatic mixed precision and gradient scaling, without gradient accumulation or clipping. Validation and testing disable gradients and Dropout.\par

\par\needspace{5\baselineskip}\begingroup\small\ttfamily\raggedright
\setlength{\parindent}{0pt}\setlength{\parskip}{1pt}
\noindent\hangindent=2em\hangafter=1\hspace*{0.0em}Algorithm S3  TRAIN\_INTERNAL\par
\noindent\hangindent=2em\hangafter=1\hspace*{0.0em}Read fixed manifests: 93 training,\allowbreak{} 23 validation,\allowbreak{} 29 test images\par
\noindent\hangindent=2em\hangafter=1\hspace*{0.0em}Read original masks for both classes\par
\noindent\hangindent=2em\hangafter=1\hspace*{0.0em}Check that growth batches do not cross partitions\par
\noindent\hangindent=2em\hangafter=1\hspace*{0.0em}Keep test images out of training and validation\par
\noindent\hangindent=2em\hangafter=1\hspace*{0.0em}for seed in [42,\allowbreak{} 7,\allowbreak{} 123,\allowbreak{} 2026]:\par
\noindent\hangindent=2em\hangafter=1\hspace*{2.0em}Set Python,\allowbreak{} NumPy,\allowbreak{} PyTorch,\allowbreak{} and CUDA random seeds\par
\noindent\hangindent=2em\hangafter=1\hspace*{2.0em}M = randomly initialized base network (S2); load no prior weights\par
\noindent\hangindent=2em\hangafter=1\hspace*{2.0em}optimizer = AdamW(M,\allowbreak{} lr=8e-5,\allowbreak{} weight\_decay=1e-4)\par
\noindent\hangindent=2em\hangafter=1\hspace*{2.0em}scheduler = ReduceLROnPlateau(mode=min,\allowbreak{} factor=0.5,\allowbreak{}\par
\noindent\hangindent=2em\hangafter=1\hspace*{16.5em}patience=8,\allowbreak{} min\_lr=2e-5)\par
\noindent\hangindent=2em\hangafter=1\hspace*{2.0em}views = 744 views from 93 images via S1 and synchronized D4\par
\noindent\hangindent=2em\hangafter=1\hspace*{2.0em}bestF1 = -infinity; no\_improvement = 0\par
\noindent\hangindent=2em\hangafter=1\hspace*{2.0em}for epoch = 1..80:\par
\noindent\hangindent=2em\hangafter=1\hspace*{4.0em}Set M to training mode; shuffle views\par
\noindent\hangindent=2em\hangafter=1\hspace*{4.0em}for (X,\allowbreak{}Ys,\allowbreak{}Yk) in views:             \# one view per update; full pass\par
\noindent\hangindent=2em\hangafter=1\hspace*{6.0em}zu,\allowbreak{} zs,\allowbreak{} zk = FORWARD(X,\allowbreak{} public\_auxiliary=false,\allowbreak{} training=true)\par
\noindent\hangindent=2em\hangafter=1\hspace*{6.0em}Yu = max(Ys,\allowbreak{} Yk)\par
\noindent\hangindent=2em\hangafter=1\hspace*{6.0em}L = BCE(zu,\allowbreak{}Yu) + (BCE(zs,\allowbreak{}Ys)+BCE(zk,\allowbreak{}Yk))/2\par
\noindent\hangindent=2em\hangafter=1\hspace*{8.0em}+ 0.2*DiceLoss(zk,\allowbreak{}Yk)\par
\noindent\hangindent=2em\hangafter=1\hspace*{6.0em}Zero gradients; backpropagate L; update optimizer once\par
\noindent\hangindent=2em\hangafter=1\hspace*{4.0em}After all training views,\allowbreak{} set M to evaluation mode\par
\noindent\hangindent=2em\hangafter=1\hspace*{4.0em}Compute mean L on 23 original-orientation validation images\par
\noindent\hangindent=2em\hangafter=1\hspace*{4.0em}Validation mask = sigmoid(zk) >= 0.50; sum Kikuchi TP/FP/FN\par
\noindent\hangindent=2em\hangafter=1\hspace*{4.0em}F1 = 2*TP / (2*TP+FP+FN)\par
\noindent\hangindent=2em\hangafter=1\hspace*{4.0em}scheduler.step(mean validation L)\par
\noindent\hangindent=2em\hangafter=1\hspace*{4.0em}Log loss,\allowbreak{} per-class metrics,\allowbreak{} learning rate,\allowbreak{} and epoch\par
\noindent\hangindent=2em\hangafter=1\hspace*{4.0em}if F1 > bestF1:\par
\noindent\hangindent=2em\hangafter=1\hspace*{6.0em}Save M; bestF1 = F1; no\_improvement = 0\par
\noindent\hangindent=2em\hangafter=1\hspace*{4.0em}else:\par
\noindent\hangindent=2em\hangafter=1\hspace*{6.0em}no\_improvement += 1\par
\noindent\hangindent=2em\hangafter=1\hspace*{4.0em}if no\_improvement >= 25: break\par
\noindent\hangindent=2em\hangafter=1\hspace*{0.0em}After all four runs,\allowbreak{} load each saved model\par
\noindent\hangindent=2em\hangafter=1\hspace*{0.0em}Evaluate 29 test images using S5 with threshold 0.45\par
\endgroup\par\medskip

Validation loss is not a secondary criterion for tied checkpoints. The learning-rate scheduler uses PyTorch's default relative improvement threshold \(10^{-4}\). AdamW momentum coefficients are (0.9, 0.999), with stabilizing constant \(10^{-8}\). The entry enables cuDNN automatic algorithm selection; fixed seeds do not guarantee bitwise identity across software and hardware.\par

\subsubsection*{S12.4 Adaptation with Nested Public Training Sets}

\addcontentsline{toc}{subsubsection}{S12.4 Adaptation with Nested Public Training Sets}

Algorithm S4 gives the complete execution order for Section 3.3. Current entry settings override grade targets and lookup weights in earlier auxiliary scripts; reproduction should use the w and t below rather than historical defaults.\par

Smaller sets are determined as follows. Read the fixed 244 training images and build queues by sample group. Shuffle each queue with seed 20260921, then stably sort by increasing frequency of the image's WLM grade in the full training set. At each step, consider only the next image in each queue and choose the candidate minimizing \(0.55a_g/n_g+0.45a_\ell/n_\ell\), where \(a_g,a_\ell\) are the numbers already selected from that group and grade, and \(n_g,n_\ell\) their totals. Ties are broken using the same random-number generator. Repeat until all images are ordered. For reproduction, preferentially read the saved sequence of 244 names to avoid differences in queue traversal or random-number implementations. Its prefixes of 20, 40, 60, 80, 100, 120, 140, 160, 180, 200, 220, and 244 images define the twelve training sets.\par

\par\needspace{5\baselineskip}\begingroup\small\ttfamily\raggedright
\setlength{\parindent}{0pt}\setlength{\parskip}{1pt}
\noindent\hangindent=2em\hangafter=1\hspace*{0.0em}Algorithm S4  ADAPT\_PUBLIC\par
\noindent\hangindent=2em\hangafter=1\hspace*{0.0em}Read 381 eligible images: 244 training,\allowbreak{} 51 validation,\allowbreak{} 86 test\par
\noindent\hangindent=2em\hangafter=1\hspace*{0.0em}Use WLM grades 5/\allowbreak{}4/\allowbreak{}3/\allowbreak{}2/\allowbreak{}1/-1; retain original grade identity\par
\noindent\hangindent=2em\hangafter=1\hspace*{0.0em}Read fixed nested training order; keep validation and test manifests fixed\par
\noindent\hangindent=2em\hangafter=1\hspace*{0.0em}w = \{5:1.00,\allowbreak{} 4:0.95,\allowbreak{} 3:0.70,\allowbreak{} 2:0.40,\allowbreak{} 1:1.00,\allowbreak{} -1:2.00\}\par
\noindent\hangindent=2em\hangafter=1\hspace*{0.0em}t = \{5:5,\allowbreak{} 4:4,\allowbreak{} 3:3,\allowbreak{} 2:2,\allowbreak{} 1:1,\allowbreak{} -1:1\}\par
\noindent\hangindent=2em\hangafter=1\hspace*{0.0em}for n in [20,\allowbreak{} 40,\allowbreak{} 60,\allowbreak{} 80,\allowbreak{} 100,\allowbreak{} 120,\allowbreak{} 140,\allowbreak{} 160,\allowbreak{} 180,\allowbreak{} 200,\allowbreak{} 220,\allowbreak{} 244]:\par
\noindent\hangindent=2em\hangafter=1\hspace*{2.0em}train = first n images; views = all 8*n D4 views after S1\par
\noindent\hangindent=2em\hangafter=1\hspace*{2.0em}for seed in [20260931,\allowbreak{} 20260932,\allowbreak{} 20260933]:\par
\noindent\hangindent=2em\hangafter=1\hspace*{4.0em}Reset global seeds; construct a new network with public auxiliary head\par
\noindent\hangindent=2em\hangafter=1\hspace*{4.0em}Load the same integrated-laboratory segmentation weights\par
\noindent\hangindent=2em\hangafter=1\hspace*{4.0em}Keep the regression head randomly initialized\par
\noindent\hangindent=2em\hangafter=1\hspace*{4.0em}optimizer = AdamW(all parameters,\allowbreak{} lr=1e-5,\allowbreak{} weight\_decay=1e-5)\par
\noindent\hangindent=2em\hangafter=1\hspace*{4.0em}Shuffle view indices with a separate Random(seed+1000)\par
\noindent\hangindent=2em\hangafter=1\hspace*{4.0em}Reshuffle after each complete pass\par
\noindent\hangindent=2em\hangafter=1\hspace*{4.0em}bestIoU = -infinity\par
\noindent\hangindent=2em\hangafter=1\hspace*{4.0em}for update = 1..24000:\par
\noindent\hangindent=2em\hangafter=1\hspace*{6.0em}Take the next view and its WLM grade ell; enter training mode\par
\noindent\hangindent=2em\hangafter=1\hspace*{6.0em}zu,\allowbreak{} zs,\allowbreak{} zk,\allowbreak{} q = FORWARD(X,\allowbreak{} public\_auxiliary=true,\allowbreak{} training=true)\par
\noindent\hangindent=2em\hangafter=1\hspace*{6.0em}Yu = max(Ys,\allowbreak{}Yk)\par
\noindent\hangindent=2em\hangafter=1\hspace*{6.0em}L = w[ell]*(BCE(zu,\allowbreak{}Yu)+BCE(zk,\allowbreak{}Yk)+0.2*DiceLoss(zk,\allowbreak{}Yk))\par
\noindent\hangindent=2em\hangafter=1\hspace*{8.0em}+ 0.35*BCE(zs,\allowbreak{}Ys) + 0.15*SmoothL1(q,\allowbreak{}t[ell])\par
\noindent\hangindent=2em\hangafter=1\hspace*{6.0em}Zero gradients; backpropagate L; update optimizer once\par
\noindent\hangindent=2em\hangafter=1\hspace*{6.0em}if update is a multiple of 400 or the final update:\par
\noindent\hangindent=2em\hangafter=1\hspace*{8.0em}Enter evaluation mode\par
\noindent\hangindent=2em\hangafter=1\hspace*{8.0em}Evaluate 51 original-orientation validation images with S5\par
\noindent\hangindent=2em\hangafter=1\hspace*{8.0em}Sum Kikuchi TP/FP/FN only for WLM grades 2/\allowbreak{}3/\allowbreak{}4/\allowbreak{}5\par
\noindent\hangindent=2em\hangafter=1\hspace*{8.0em}J = TP/(TP+FP+FN)\par
\noindent\hangindent=2em\hangafter=1\hspace*{8.0em}if J > bestIoU: save model; bestIoU = J\par
\noindent\hangindent=2em\hangafter=1\hspace*{8.0em}Log update,\allowbreak{} loss,\allowbreak{} and validation statistics\par
\noindent\hangindent=2em\hangafter=1\hspace*{4.0em}Load the saved model; evaluate the fixed 86 test images with S5\par
\noindent\hangindent=2em\hangafter=1\hspace*{4.0em}Store per-image counts\par
\noindent\hangindent=2em\hangafter=1\hspace*{0.0em}Evaluate the same starting segmentation model directly on 86 test images\par
\noindent\hangindent=2em\hangafter=1\hspace*{0.0em}This gives the result with zero adaptation images\par
\noindent\hangindent=2em\hangafter=1\hspace*{0.0em}Join all test records to three-reader ratings using S6 for final reporting\par
\endgroup\par\medskip

SmoothL1 uses the transition point 1 defined in Section 3.3. The generic \texttt{weak\_weight} fallback does not override the six-grade lookup weights. Tied validation scores retain the earlier checkpoint.\par

Zero adaptation images means evaluating only the starting segmentation model, without interpreting an untrained regression head. Presentation controls in S5 reuse this algorithm with 9,840 updates for 100 images and 58,560 for 244, leaving other settings unchanged. Validation also occurs at the endpoint when it is not divisible by 400.\par

\subsubsection*{S12.5 Single-Image Inference and Region Counts}

\addcontentsline{toc}{subsubsection}{S12.5 Single-Image Inference and Region Counts}

The following procedure produces both regions and specifies Kikuchi metric calculation. Automated statistics for the primary public experiment actually store only per-image Kikuchi counts; streak visualizations use the same class head and geometry rules. Valid original-image extent is determined from geometry, not by whether a pixel is black, so genuinely black regions within an original still enter evaluation.\par

\par\needspace{5\baselineskip}\begingroup\small\ttfamily\raggedright
\setlength{\parindent}{0pt}\setlength{\parskip}{1pt}
\noindent\hangindent=2em\hangafter=1\hspace*{0.0em}Algorithm S5  INFER\_AND\_COUNT(saved\_model,\allowbreak{} record,\allowbreak{} experiment,\allowbreak{} tau)\par
\noindent\hangindent=2em\hangafter=1\hspace*{0.0em}Set evaluation mode; disable gradients and Dropout\par
\noindent\hangindent=2em\hangafter=1\hspace*{0.0em}X,\allowbreak{} Ys,\allowbreak{} Yk,\allowbreak{} Gs,\allowbreak{} Gk,\allowbreak{} geometry = PREPARE(record,\allowbreak{} experiment)\par
\noindent\hangindent=2em\hangafter=1\hspace*{0.0em}zu,\allowbreak{} zs,\allowbreak{} zk,\allowbreak{} [q] = FORWARD(X,\allowbreak{} auxiliary head if present,\allowbreak{} training=false)\par
\noindent\hangindent=2em\hangafter=1\hspace*{0.0em}Ps256 = sigmoid(zs) >= tau\par
\noindent\hangindent=2em\hangafter=1\hspace*{0.0em}Pk256 = sigmoid(zk) >= tau\par
\noindent\hangindent=2em\hangafter=1\hspace*{0.0em}if experiment == public:\par
\noindent\hangindent=2em\hangafter=1\hspace*{2.0em}Ps = Resize(RemovePadding(Ps256),\allowbreak{} original dimensions,\allowbreak{} nearest)\par
\noindent\hangindent=2em\hangafter=1\hspace*{2.0em}Pk = Resize(RemovePadding(Pk256),\allowbreak{} original dimensions,\allowbreak{} nearest)\par
\noindent\hangindent=2em\hangafter=1\hspace*{2.0em}references = original-size Gs/Gk\par
\noindent\hangindent=2em\hangafter=1\hspace*{0.0em}else:\par
\noindent\hangindent=2em\hangafter=1\hspace*{2.0em}Ps,\allowbreak{} Pk = Ps256,\allowbreak{} Pk256\par
\noindent\hangindent=2em\hangafter=1\hspace*{2.0em}references = Ys/Yk on the 256 grid\par
\noindent\hangindent=2em\hangafter=1\hspace*{0.0em}for (P,\allowbreak{}G) in [(Ps,\allowbreak{}streak reference),\allowbreak{} (Pk,\allowbreak{}Kikuchi reference)]:\par
\noindent\hangindent=2em\hangafter=1\hspace*{2.0em}TP = Count(P AND G);  FP = Count(P AND NOT G)\par
\noindent\hangindent=2em\hangafter=1\hspace*{2.0em}FN = Count(NOT P AND G);  TN = Count(NOT P AND NOT G)\par
\noindent\hangindent=2em\hangafter=1\hspace*{2.0em}Store original name,\allowbreak{} source group,\allowbreak{} reference identity,\allowbreak{} threshold,\allowbreak{}\par
\noindent\hangindent=2em\hangafter=1\hspace*{5.0em}TP/FP/FN/TN\par
\noindent\hangindent=2em\hangafter=1\hspace*{0.0em}return Ps,\allowbreak{} Pk,\allowbreak{} class counts,\allowbreak{} [continuous grade q]\par
\endgroup\par\medskip

Thresholds are specified in Algorithms S3 and S4. Mean-score ensemble displays follow their captions and do not replace primary statistics. To reproduce diagnostic scores for the auxiliary union head, separately evaluate sigmoid(zu) against \(G_s\lor G_k\), rather than substituting the logical OR of the two class predictions.\par

\subsubsection*{S12.6 Pooling, Three-Reader Strata, and Uncertainty}

\addcontentsline{toc}{subsubsection}{S12.6 Pooling, Three-Reader Strata, and Uncertainty}

Reporting uses counts directly and assesses visibility separately from reference nonemptiness. Predictions on empty references remain in the statistics. Historical per-image CSVs used zero as a placeholder for empty–empty IoU; here these cases are recognized from counts as NA, not zero scores in image averages.\par

\par\needspace{5\baselineskip}\begingroup\small\ttfamily\raggedright
\setlength{\parindent}{0pt}\setlength{\parskip}{1pt}
\noindent\hangindent=2em\hangafter=1\hspace*{0.0em}Algorithm S6  REPORT(frozen\_counts,\allowbreak{} human\_ratings)\par
\noindent\hangindent=2em\hangafter=1\hspace*{0.0em}for each model and each test image:\par
\noindent\hangindent=2em\hangafter=1\hspace*{2.0em}d = TP+FP+FN\par
\noindent\hangindent=2em\hangafter=1\hspace*{2.0em}imageIoU = TP/d if d>0 else NA\par
\noindent\hangindent=2em\hangafter=1\hspace*{2.0em}empty\_reference = (TP+FN == 0)\par
\noindent\hangindent=2em\hangafter=1\hspace*{2.0em}If empty,\allowbreak{} also record FP,\allowbreak{} valid pixel count,\allowbreak{} and prediction nonemptiness\par
\noindent\hangindent=2em\hangafter=1\hspace*{0.0em}for each public test image:\par
\noindent\hangindent=2em\hangafter=1\hspace*{2.0em}grades = [WLM grade,\allowbreak{} HHS grade,\allowbreak{} ZZH grade]\par
\noindent\hangindent=2em\hangafter=1\hspace*{2.0em}Map -1 to 1; median = middle grade after sorting\par
\noindent\hangindent=2em\hangafter=1\hspace*{2.0em}stratum = 4-5 clear / 3 discernible / 2 ambiguous / 1 no target\par
\noindent\hangindent=2em\hangafter=1\hspace*{0.0em}for each training size,\allowbreak{} model,\allowbreak{} and reported set (all or a stratum):\par
\noindent\hangindent=2em\hangafter=1\hspace*{2.0em}Sum TP,\allowbreak{} FP,\allowbreak{} FN,\allowbreak{} TN separately over images in the set\par
\noindent\hangindent=2em\hangafter=1\hspace*{2.0em}IoU = sumTP/(sumTP+sumFP+sumFN); NA if denominator is zero\par
\noindent\hangindent=2em\hangafter=1\hspace*{2.0em}F1 = 2*sumTP/(2*sumTP+sumFP+sumFN); NA if denominator is zero\par
\noindent\hangindent=2em\hangafter=1\hspace*{2.0em}For no-target images,\allowbreak{} FP area fraction = sumFP/total valid pixels\par
\noindent\hangindent=2em\hangafter=1\hspace*{0.0em}for each training size:\par
\noindent\hangindent=2em\hangafter=1\hspace*{2.0em}Report mean and sample standard deviation (ddof=1) across models\par
\noindent\hangindent=2em\hangafter=1\hspace*{2.0em}Zero adaptation images has one evaluation; report no training SD\par
\noindent\hangindent=2em\hangafter=1\hspace*{0.0em}Save per-image counts,\allowbreak{} rating joins,\allowbreak{} per-model results,\allowbreak{} and summaries\par
\noindent\hangindent=2em\hangafter=1\hspace*{0.0em}Do not replace them with a single best model\par
\endgroup\par\medskip

For equally weighted internal image results, first average IoU across the four models for each image, then average over 29 images. For equally weighted batch results, further average images within each batch and finally average the six batches. The two 95\% intervals resample the 29 image means or six batch means with replacement 20,000 times and take the 2.5th and 97.5th percentiles. Equal-batch weighting does not first pool pixels within batches to compute IoU.\par

Public overall-IoU uncertainty uses 5,000 draws of the 12 test sample groups, drawing 12 groups with replacement each time and retaining all images and multiplicities from each sampled group. Pooled pixel-level IoU is recalculated for each model, then averaged across models. Comparisons of training sizes use the same sampled groups and paired seeds on both sides. Confidence intervals use the 2.5th and 97.5th percentiles of sampled differences or means. Error bars in stratified tables and curves are sample standard deviations across training runs, distinct from these group-resampling intervals. The six internal design controls use 100,000 draws as specified in S8.\par
\end{document}